\documentclass[manuscript]{aastex}

\usepackage{natbib}
\usepackage{color}

\newcommand{\beq}{\begin{equation}}
\newcommand{\eeq}{\end{equation}}
\newcommand{\beqa}{\begin{eqnarray}}
\newcommand{\eeqa}{\end{eqnarray}}

\slugcomment{The Astrophysical Journal}

\shorttitle{Fast Magnetic Twister in a Coronal Arcade}
\shortauthors{Murawski et al.}

\begin{document}


\title{Fast Magnetic Twister and Plasma Perturbations in a 3-D Coronal Arcade}


\author{K. Murawski}
\affil{Group of Astrophysics, UMCS, ul. Radziszewskiego 10, 20-031 Lublin, Poland}
\email{kmur@kft.umcs.lublin.pl}

\author{A. K. Srivastava}
\affil{Department of Physics, Indian Institute of Technology (BHU), Varanasi-221005, India}
\email{asrivastava.app@iitbhu.ac.in}

\author{Z. E. Musielak}
\affil{Department of Physics, University of Texas at Arlington, Arlington, TX 76019, 
USA}
\email{zmusielak@uta.edu}


%
\begin{abstract}
We present results of 3-D numerical simulations of a fast magnetic twister 
excited above a foot-point of the potential solar coronal arcade that is 
embedded in the solar atmosphere with the initial VAL-IIIC temperature 
profile, which is smoothly extended into the solar corona. With the use of 
the FLASH code, we solve 3-D ideal magnetohydrodynamic equations by 
specifying a twist in the azimuthal component of magnetic field in the 
solar chromosphere. The imposed perturbation generates torsional Alfv\'en 
waves as well as plasma swirls that reach the other foot-point of the 
arcade and partially reflect back from the transition region. 
The two vortex channels are evident in the generated twisted flux-tube with a fragmentation 
near its apex that results from the initial twist as well as from the 
morphology of the tube. The numerical results are compared to observational 
data of plasma motions in a solar prominence. 
The comparison shows that the numerical results and the data qualitatively agree even though the observed 
plasma motions occur over comparatively large spatio-temporal scales in the prominence. 
\end{abstract}


\section{Introduction}
Vortex and swirling plasma motions are ubiquitous in the solar atmosphere. 
At small spatio-temporal scales the chromospheric swirling motions were discovered 
as super-tornadoes providing an alternative mechanism for channeling energy 
up to the inner solar corona (Wedemeyer-B{\"o}hm et al. 2012, 2013). Just 
after this novel observation, Su et al. (2013) reported the large-scale but 
slowly rotating coronal tornadoes. They studied the rotating vertical magnetic 
structures most likely driven by the underlying vortex flows in the photosphere 
that existed in a group with prominences. In case of such large-scale and 
slowly rotating tornadoes, which are long living magnetic structures, several 
other observational findings were made using recent space-borne observations 
(e.g., Panesar et al. 2013; Yan et al. 2013; Wedemeyer-B{\"o}hm et al. 2013; 
and references therein). However, recently the findings of Panasenco et al. (2014) 
poses a challenge to the large-scale coronal tornado detections. 
In the observational data they analyzed and found that the coronal tornado-like 
appearance usually associated with the prominences is mainly an illusion due 
to projection effects.

The fact that the projection effects can significantly distort the 
interpretation of such observations of coronal tornadoes is known. Similar
effects could become important in interpretation of the solar observations 
of various localized plasma motions, such as plasma cyclones, swirls and 
tornadoes, which are ubiquitous in the solar corona, and whose origin, 
nature, morphology, life-time and driving mechanisms are difficult to 
observationally determine. Our numerical studies of magnetic twisters 
in solar coronal arcades presented in this paper are designed to address 
some of these currently unsolved problems. 

The exact physical mechanisms responsible for the origin of long-lived 
coronal tornadoes are not fully known. Moreover, the relationship between 
the short-lived chromospheric tornadoes and the long-lived coronal tornadoes is 
also not well established. Shukla (2013) developed a generalized theory, which 
implies that the modified-kinetic Alfv\'en waves (m-KAWs) in a magnetized plasma 
can propagate in the form of tornadoes, characterized by the plasma whirls or 
magnetic flux ropes carrying orbital angular momentum. Now, in the lower solar 
atmosphere, the small-scale chromospheric tornadoes may be caused by the photospheric 
vortices (e.g., Bonet et al. 2008; Shelyag et al. 2011; Murawski et al. 2013a,b, 
and references therein). Moreover, various types of these vortex motions could be 
associated with the different eddies and waves depending on the localized 
plasma and magnetic conditions as well as on the nature of the drivers/perturbations
(Fedun et al. 2011; Murawski et al. 2013b). 
Therefore, it is important to understand
the nature and generation of such swirling plasma motions in 
the different layers of the solar atmosphere, where various topologies of solar magnetic fields occur. 

As far as the large-scale solar tornadoes are concerned, they are mostly observed in 
the association with the prominence legs (Su et al. 2012; Wedemeyer-B{\"o}hm et al. 
2013), and are most likely related to the response of the photospheric vortices 
activated near the foot-points. These vortex flows exhibit spiral motions 
both upwards and downwards with speeds comparable to the values found for such 
tornado rotations in the form of prominence barbs (Wedemeyer-B{\"o}hm et al. 2012, 
Su et al. 2012). However, they are long-lived (on the time-scale of $12-24$ hours) 
but slow rotators. There is another kind of interesting twisting and then 
swirl of the plasma around the core prominence magnetic field that may have the same 
origin due to rotation near the prominences' foot-point but entirely different 
when compared to the long-lived slow tornadoes. We call it a fast magnetic 
twister and associate it with a coronal arcade. This paper is devoted to numerical
studies of such a phenomenon. 

It must be pointed out that the fast magnetic twisters, which live on the 
time-scale of minutes, are seldomly observed in the solar atmosphere. 
Recently, Joshi et al. (2014) investigated a fast twisting prominence system in the context 
of its stability and reformation that is significant for space-weather prediction. 
However, their aim was certainly not to understand the possible driver responsible 
for the evolution of fast magnetic twister and plasma swirl in the filament system 
that was initially bipolar in nature. 
It should also be noted that Shelyag et al. (2013) studied solar photospheric vortices 
by using MHD simulations with non-gray 
radiative transport and a non-ideal equation of state. The main difference between 
these two studies is that only the former takes into account variation of the local 
velocity field on time. As a result, the photospheric vortices (tornado-like 
motions) do not exist but instead torsional Alfv\'en waves are generated. 
We discuss this finding after we present the results of our numerical simulations.   

In the present paper, we perform 3-D numerical simulations of the evolution 
of right-handed clock-wise twist and its responses in form of the torsional 
Alfv\'en and plasma perturbations in an initial potential field configuration of 
a magnetic flux-tube mimicking the observed prominence system. 
We call the magneto-plasma motions that evolved in our model "fast magnetic twisters".
Since 3-D numerical simulations in realistic atmosphere would be extremely 
computationally expensive to model the observed prominence flux-tube and its associated dynamics, 
we rescale the realistic very large spatio-temporal
scales characteristic for the observational domain to a much smaller 3-D 
numerical simulation domain. Nevertheless, our simulation results match 
qualitatively the observed magnetic field and plasma dynamics as it 
is shown in our comparison between the numerical results and some recent solar observations. 

This paper is organized as follows. In Sec.~\ref{sec:num_model}, we 
present the numerical simulation model. We outline the results of 
numerical simulations in the context of its physical significance 
in Sec.~3. In Sec.~4, we compare the numerical results with the 
relevant solar observational data. In the last section, we 
present discussion and concluding remarks.
\section{Numerical Model and Results of Numerical Simulations}
\label{sec:num_model}
\subsection{Governing Equations}
We consider the following set of ideal MHD equations 
\beqa
\label{eq:MHD_rho}
{{\partial \varrho}\over {\partial t}}+\nabla \cdot (\varrho{\bf V})=0\, ,\\
\label{eq:MHD_V}
\varrho{{\partial {\bf V}}\over {\partial t}}+ \varrho\left ({\bf V}\cdot 
\nabla\right ){\bf V} =
-\nabla p+ \frac{1}{\mu}(\nabla\times{\bf B})\times{\bf B} +\varrho{\bf g}\, ,\\
\label{eq:MHD_p}
{\partial p\over \partial t} + \nabla\cdot (p{\bf V}) = (1-\gamma)p \nabla 
\cdot {\bf V}\, ,\\
\label{eq:MHD_B}
{{\partial {\bf B}}\over {\partial t}}= \nabla \times ({\bf V}\times{\bf B})\, , 
\hspace{3mm}
\nabla\cdot{\bf B} = 0\, ,
\eeqa
where ${\varrho}$ denotes mass density, ${\bf V}=[V_{\rm x},V_{\rm y},V_{\rm z}]$ 
is the flow velocity, ${\bf B}=[B_{\rm x},B_{\rm y},B_{\rm z}]$ is the magnetic field, 
$p$ is the gas pressure, $T$ is the temperature, $\mu$ represents the magnetic permeability, 
$\gamma=5/3$ is the adiabatic index, 
${\bf g}=(0,-g,0)$ is the gravitational acceleration with its magnitude $g=274$ m s$^{-2}$ being the solar value, 
$m$ denotes mean particle mass that is specified by mean molecular weight value of $0.6$, 
and $k_{\rm B}$ is the Boltzmann's constant. 
\subsection{Equilibrium Configuration}
We assume that the above set of MHD equations describes the solar atmosphere,
which is initially in static equilibrium 
(${\bf V}_{\rm e}={\bf 0}$) with the pressure gradient balanced by the gravity force
\begin{equation}
\label{eq:p}
-\nabla p_{\rm e} + \varrho_{\rm e} {\bf g} = {\bf 0}\, ,
\end{equation}
and with $\varrho_{\rm e}$ and $p_{\rm e}$ being the equilibrium mass density and gas pressure, respectively. 
Using the ideal gas law and the $y$-component of Eq.~(\ref{eq:p}), we obtain 
\begin{equation}
\label{eq:pres}
p_{\rm e}(y)=p_{\rm 0}~{\rm exp}\left[ -\int_{y_{\rm r}}^{y}\frac{dy^{'}}
{\Lambda (y^{'})} \right]\, ,\hspace{3mm}
\label{eq:eq_rho}
\varrho_{\rm e} (y)=\frac{p_{\rm e}(y)}{g \Lambda(y)}\, ,
\end{equation}
where
\begin{equation}
\Lambda(y) = k_{\rm B} T_{\rm e}(y)/(mg)
\end{equation}
is the pressure scale-height, and $p_{\rm 0}$ denotes the gas 
pressure at $y=y_{\rm r}$. 

\begin{figure}
\begin{center}
\includegraphics[scale=0.45, angle=0]{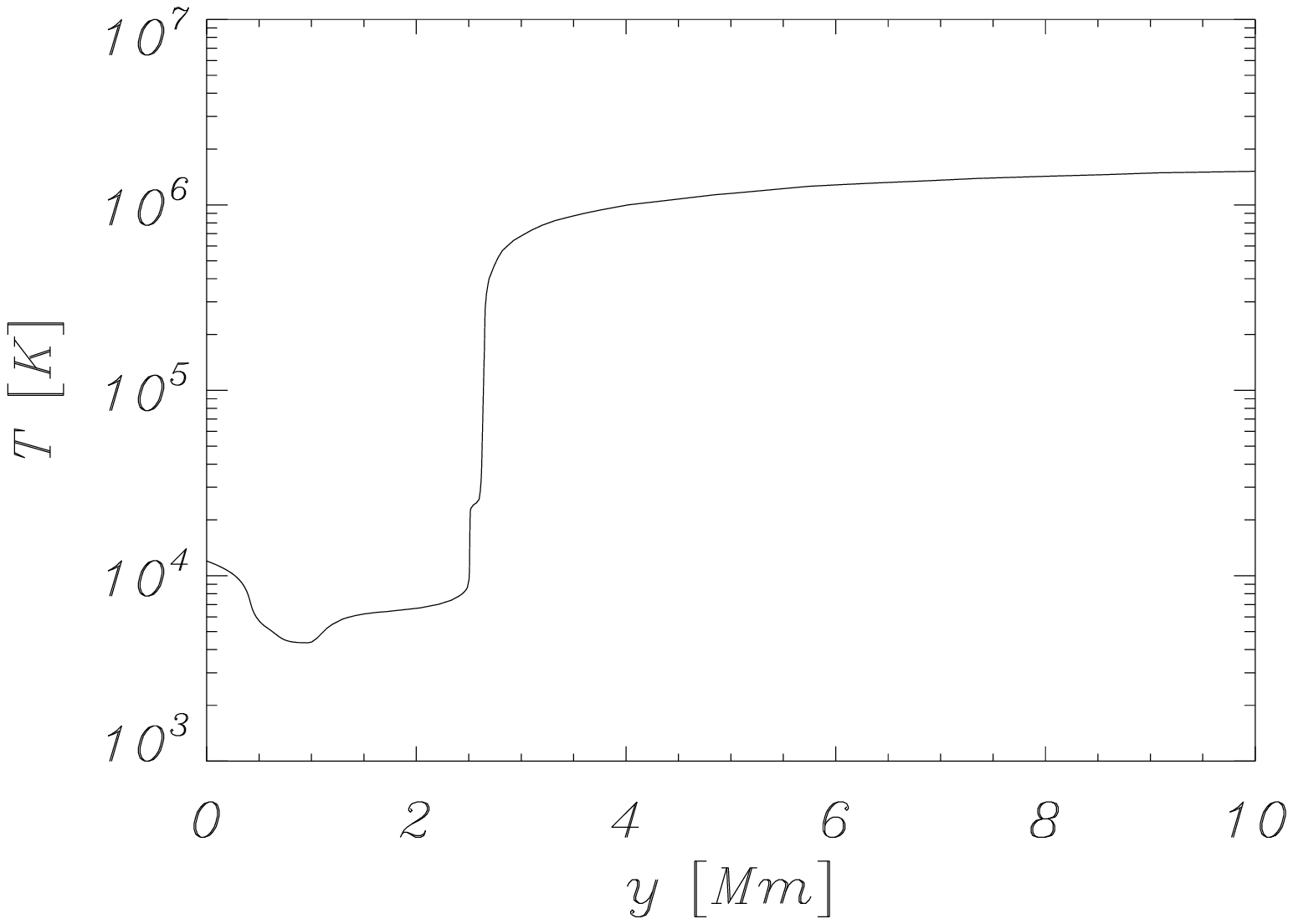}
\includegraphics[scale=0.35, angle=0]{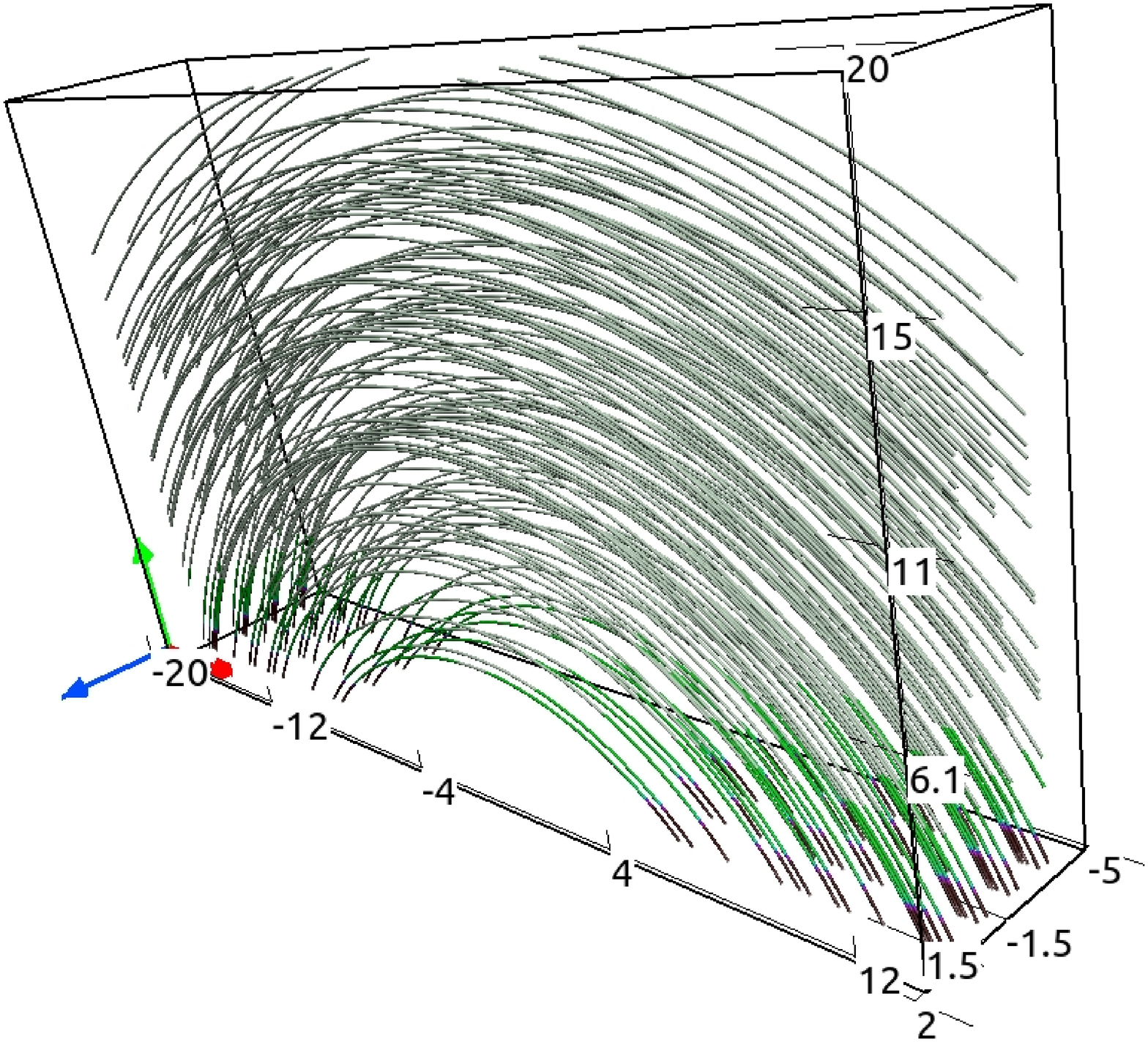}
\vspace{-2.0cm}
\caption{
Equilibrium profiles of temperature (top panel) and 
magnetic field lines (bottom panel). 
The 3D visualization of 
the simulation data is carried out using the VAPOR (Visualization and Analysis Platform) software package. 
}
\label{fig:initial_profile}
\end{center}
\end{figure}

We assume that the equilibrium temperature profile $T_{\rm e}(y)$ 
of the solar atmosphere is given by the VAL-C (Vernazza et al. 1981) 
atmospheric model, which is smoothly extended into the solar 
corona (Fig.~\ref{fig:initial_profile}, the top panel). Then with 
Eq.~(\ref{eq:pres}), we obtain the corresponding gas pressure and 
mass density profiles (not shown). In this model the temperature 
reaches $1$ MK at coronal heights and saturates at this level. The 
atmosphere is structured so that the solar photosphere occupies the 
region $0 < y < 0.5$ Mm, the solar chromosphere is sandwiched between 
$y=0.5$ Mm and the transition region that is located at $y\simeq 2.7$ 
Mm. The temperature minimum is acquired at $y\simeq 0.9$ Mm 
(Fig.~\ref{fig:initial_profile}, top).

As a result of Eq.~(\ref{eq:p}) a magnetic field must be force-free 
and the required condition is 
\begin{equation}\label{eq:B_e}
(\nabla\times{\bf B}_{\rm e})\times{\bf B}_{\rm e} = {\bf 0}\ , 
\end{equation}
such that it satisfies the current free condition 
\beq
\nabla \times {\bf B}_{\rm e}={\bf 0}\, ,
\eeq
and it is specified by the magnetic flux function, $A(x,y)$, defined by
\begin{equation}\label{eq:B_e1}
{\bf B}_{\rm e}=\nabla \times (A\hat {\bf z})\, ,
\end{equation}
where the subscript $_{\rm e}$ corresponds to equilibrium quantities 
and $\hat {\bf z}$ is a unit vector along $z$-direction. 

We set an arcade magnetic field by choosing
\begin{equation}
A(x,y) = \frac{b}{2}B_{\rm 0} \ln [{x^2+(y-b)^2}]\, 
\end{equation}
with $B_{\rm 0}$ being the magnetic field at the reference level 
and $b$ being the vertical coordinate of the singularity. We set 
and hold fixed $b=-5$ Mm while $B_{\rm 0}$ is determined from the 
following condition: 
\begin{equation}
c_{\rm A}(x=0,y=y_{\rm r}) = 10 c_{\rm s}(y=y_{\rm r})\, ,
\end{equation}
where the Alfv\'en, $c_{\rm A}$, and sound, $c_{\rm s}$, speeds are 
given by 
\beq
c_{\rm A}(x,y) = \sqrt{\frac{B_{\rm e}^2(x,y)}{\mu \varrho_{\rm e}(y)}}\, ,
\hspace{3mm} 
c_{\rm s}(y) = \sqrt{\frac{\gamma p_{\rm e}(y)}{\varrho_{\rm e}(y)}}\, ,
\eeq
and $y_{\rm r} = 10$ Mm is the reference level that we choose in the solar 
corona at $y_{\rm r}=10$ Mm. The corresponding magnetic field-lines are 
displayed in Fig.~\ref{fig:initial_profile} (the bottom panel). 
\subsection{Results of Numerical Simulation}
\begin{figure}
\begin{center}
\includegraphics[scale=0.5, angle=90]{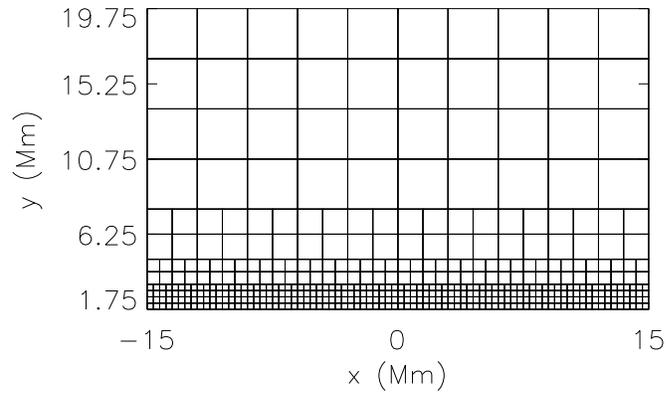}
\vspace{-2.0cm}
\caption{ Numerical blocks with their boundaries (solid lines) 
at $t=0$ s in the vertical plane ($z=0$). 
}
\label{fig:blocks}
\end{center}
\end{figure}

We solved the set of MHD equations (\ref{eq:MHD_rho})-(\ref{eq:MHD_B}) 
numerically using the FLASH code (Lee and Deane 2009, Lee 2013). This 
code implements a second-order unsplit Godunov solver with various slope 
limiters and Riemann solvers. We set the simulation box as $(-15,15)$ Mm 
$\times$ $(1.75,19.75)$ Mm $\times$ $(-3, 3)$ Mm. At the top, bottom, 
left- and right-hand sides of the numerical domain we imposed the boundary 
conditions by fixing in time all plasma quantities to their equilibrium 
values, while along $z$ we implemented open boundaries. Additionally, 
at the bottom boundary a twist in the azimuthal component of magnetic field, 
$B_{\theta}$, was specified using
\beqa\label{eq:perturb}
B_{\theta} (x,y,t) = - A_{\rm B}\, 
r_{\rm h}\, 
\exp\left[ 
-\frac{r_{\rm h}^2 +(y-y_{\rm 0})^2} {w_{\rm}^2}
\right]
\left[ \exp\left(\frac{t}{\tau}\right) -1 \right]
\, ,
\eeqa
where $r_{\rm h}^2 \equiv (x-x_{\rm 0})^2+(z-z_{\rm 0})^2$, 
$A_{\rm B}$ is the amplitude of the pulse, $(x_{\rm 0},y_{\rm 0},
z_{\rm 0})$ is its position, $w_{\rm }$ denotes its width, and 
$\tau$ is a growing time of the implemented twist. 
We set and hold fixed $A_{\rm B}=0.05$ Tesla, $x_{\rm 0}=-10$ Mm, 
$y_{\rm 0}= 1.75$ Mm, $z_{\rm 0}= 0$ Mm, $w_{\rm }=0.3$ Mm, and 
$\tau=100$ s. Therefore, the spatial scale of numerical domain 
is $18$ Mm in the vertical direction, $30$ Mm in the horizontal 
$x$-direction, and $6$ Mm in the horizontal $z$-direction. In 
our simulations, we use an Adaptive Mesh Refinement (AMR) grid 
with a minimum (maximum) level of refinement set to $2$ $(5)$. 
The extent of the simulation box in the $y$-direction ensures 
that we catch the essential physics occurring in the solar 
photosphere-low corona domain, and minimize the effect of 
spurious signal reflections from the top boundary. As each 
block consists of $8\times 8\times 8$ identical numerical cells, 
we reach the effective finest spatial resolution of about $0.24$ 
Mm, below the altitude $y=3.25$ Mm. The initial system of blocks 
is shown in Figure~\ref{fig:blocks}. 

Fig.~\ref{fig:B_lines_w} shows a development of the initial magnetic 
field configuration of an initially potential arcade embedded in the
corona whose foot-points were anchored in the photosphere. The magnetic 
field lines were initially parallel to each other without any twist 
present. We applied a twist in the B$_{\theta}$ component of the 
magnetic field (Eq.~\ref{eq:perturb}) above the photosphere. The 
figure displays the activation of a right-handed clockwise twist
in the magnetic field near the chromosphere into the corona. The 
snapshots respectively represent the following times: $t=50$ s, 
$t=75$ s, $t=100$ s, $t=125$ s shown from the top-left to bottom-right 
panels. Red, green, and blue arrows correspond to the $x$-, $y$-, 
and $z$-axis, respectively, 
and such notation is used in all 3-D figures throughout this paper. 
It is clear that initially the azimuthal 
component of magnetic field lines warps within the ambient potential 
field as per the right-handed twisting (top-left). 
However, at the later times, the twisting scenario becomes more 
complex and it does not remain in the state of ideal right-handed 
circular twist. Instead, the depression from one-side is seen in 
the chromospheric region that makes the evolution of eight-shape 
complex twisting of the magnetic field lines near the foot-point 
of the flux-tube. On the contrary, the right-handed twist evolves 
more ideally circularly at the coronal heights as shown in 
Fig.~\ref{fig:B_lines_w} (the top-right and bottom-left panels). The 
shearing of whole flux-tube is also evident during the activation 
of the twist and its propagation higher into the corona 
(see the bottom-right panel in Fig.~\ref{fig:B_lines_w}).

Fig.~\ref{fig:B_lines} illustrates evolution of the magnetic field 
lines in the zoomed region that is located close to the implemented 
twist in $B_{\theta}$. The initially parallel magnetic field lines 
become changed after applying a twist in the B$_{\theta}$ component 
of the magnetic field (see Eq.~\ref{eq:perturb}). This figure displays 
the evolution of a right-handed clockwise twist in the magnetic 
field of the flux-rope from its chromospheric view point in the plane
perpendicular to our line-of-sight (LOS). The snapshots respectively
represent the following times: $t=50$ s (the top-left panel), $t=75$ s 
(the top-right panel), $t=100$ s (the bottom-left panel), and $t=125$ 
s (the bottom-right panel). 
It is clear that at 
$t=50$ s the twist is applied at the left foot-point and the magnetic 
field lines start bending from left-to-right (clock-wise), which is 
the sign of right-handed positive twist (top-left). Then, this twist 
grows in the coronal heights along the equilibrium initial magnetic 
field configuration (the top-right, bottom-left, bottom-right panels). 

Fig.~\ref{fig:V-lines} shows the temporal snapshots of streamlines in 
the velocity field. These streamlines are given by  
\begin{equation}
\frac{dx}{V_{\rm x}} = \frac{dy}{V_{\rm y}} = \frac{dz}{V_{\rm z}} \, .
\end{equation} 
The figure displays the activation of a plasma swirling motion. The 
snapshots respectively represent the following times: $t=50$ s, $t=75$ s, 
$t=100$ s, and $t=125$ s as shown from the top-left to bottom-right panels. 
The plasma starts swirling and also rising up in the twisted magnetic fields. 
At the time $t=50$ s, the helical (right-handed) swirling motion is more 
evolved and it reached another foot-point at $t=100$ s (the top-right panel). 
Indeed these are fast perturbations moving inside the flux-tube from left 
foot-point to the right foot-point with a speed of about $500$ km s$^{-1}$. 
The perturbations are the torsional Alfv\'en wave-like fast perturbations 
triggered in the considered coronal arcade system. The plasma motion is
very complex as it is seen in the evolution of two vortex channels in 
the presented snapshots. Moreover, the whole arcade system whips 
in the vertical direction, which may be the effect of kink instability 
seeded by twisted magnetic field lines. Finally, after the full activation 
of the twist, the arcade system possess torsional Alfv\'en wave-like 
fast perturbations, two parallel vortex channels, and plasma swirling in them.
We shall discuss their physical consequences in the next section.

Fig.~\ref{fig:log_rho} displays the mass density maps in X-Z plane at 
the following times: $t=50$ s (the top-left panel), $t=75$ s (the 
top-right panel), $t=100$ s (the bottom-left panel), and $t=125$ s (the 
bottom-right panel). The evolution of shock front along the magnetic 
field lines and the complex plasma motions are evident in these snapshots. 
Again, the plasma motion is very complex because it is associated 
with the two vortex systems and their plasma swirling. The twisted 
magnetic field lines squeeze the chromospheric plasma and push it upwards 
(the top-left panel). Some plasma becomes detached as shown in the top-right
panel. However, the plasma finally spreads over the curved interfaces. 
This overall dynamical scenario is associated with the evolution of the
fast magnetic twister and the plasma perturbations in the large-scale 
magnetic structure. 
\section{Physical Interpretation}
The results of our numerical simulations clearly show the presence 
of shearing, complex evolution and fragmentation of the twist as well as 
the generation of two sets of vortex channels, and the excitation of the 
fast torsional Alfv\'en wave-like perturbations in both channels 
(Figs.~\ref{fig:B_lines_w}-\ref{fig:log_rho}). The possible underlying physical 
scenarios are discussed below. 

It is seen that some shearing occurs in the whole body of the evolved 
twisted magnetic flux-tube up to the corona (cf., Fig.~\ref{fig:B_lines}). 
The underlying physics reveals that the shearing motions are generated 
by the competing actions of the Lorentz force and the gravity. Thus, these
two forces are responsible for driving the two counter-rotating twisters. 
As the cross section of these twisters expands due to the stratification in 
the solar atmosphere, the mass density (or a gas pressure) stratification of 
the plasma causes the upper parts of the twister to expand while the lower 
part remains constricted in the solar chromosphere, which results in the 
shearing as evident in our numerical results. Such shearing of the 
loops were first reported by Manchester (2003) and Manchester et al. (2004). 

Depending on the initial degree of twist, portions of the magnetic 
flux-tube shed vortex pairs, and the magnetic flux is redistributed in 
the tube cross section in such a way that much of the flux is located 
away from the tube's central axis. This phenomenon is known as the 
fragmentation of the tube and it was already recognized in the emerging 
bipolar flux-tubes from the sub-photospheric layers in the outer solar 
atmosphere with some twists (e.g., Abbett et al. 2000, Srivastava et al. 2010). 
However, the first numerical evidence for the generation 
of such fragmentation is clearly given by the results of our 3-D 
simulations of the activation of twist in the coronal arcade. 
The rising bipolar magnetic flux-tubes from the solar convection zone 
up to the photosphere can undergo the fragmentation if their initial 
twist is less than a critical twist and the curvature of the apex is 
small (Abbett et al. 2000). Our numerical simulations do support this
already established physical picture.

Another interesting phenomenon seen in our 3D numerical simulations 
is that the velocity streamlines over the two vortex pairs reach the other 
foot-point of the magnetic flux-tube with an average speed of about 
$400-500$ km s$^{-1}$, which is associated with the fast moving (with 
the Alfv\'en speed) perturbation. This is the most likely evolution of 
torsional Alfv\'en perturbations generated during twisting of the magnetic 
field lines and associated with the swirl/vortex motions.

The recent work by Shelyag et al. (2013) shows that the solar photospheric 
vortices do not exist because the perturbations used by these authors 
produce torsional Alfv\'en waves instead. This is an interesting result, 
which is related to vortices in the intergranular cells that occur on smaller 
spatial and temporal scales than those considered in this paper. In our case, 
the large-scale magnetic system does show both the vortex channels as well as
torsional Alfv\'en waves.
\section{Qualitative Match of the Numerical Results with the Observations}
We now describe observational data obtained for a solar prominence and
make comparison between the data and our numerical results. Since 
there is a significant difference in spatio-temporal scales between the 
observed prominence and the numerically simulated fast magnetic twister 
in an arcade, the comparison can only be qualitative. Nevertheless, 
a qualitative agreement between the data and theory can be seen.

On 4 August, 2013 during 11:20 -12:20 UT there was an interesting plasma
motion seen near the disk-centre in the south-east quadrant of the Sun. 
An almost bipolar core filament was lying quiescently, and suddenly 
some twisting and brightening occurred at its left foot-point; the 
nature of this twist was right-handed and clockwise 
(see Fig.~\ref{fig:obser} and movie Twisted-Filament.mp4). The figure and movie
clearly show plasma swirls from the left-to-right clockwise, and fine 
structured motions of the plasma reached at another foot-point of the 
filament flux-rope system in almost $1800$ s with the average speed of 
$\sim 400$ km s$^{-1}$. It should be noted that almost semi-circular 
flux-rope system has the length $\sim 700$ Mm, while the width near its 
apex was almost $70$ Mm (cf., 11:49 UT snapshot) in the projection. The 
average speed of the moving perturbations from the left to right foot-point 
is indeed a fast speed. The plasma motions show some fragmentation of the 
path (see the 11:49 UT snapshot) and they convert into 
a very complex interaction between the two fragmented branches; as a result of
this interaction a shallow apex is created with an apparent dip (see the 
11:59 UT snapshot). The opposite plasma motions are also evident 
in the remaining time span up to some extent, while the main 
right-handed vortex or swirl motions still continue for the next 
30 minutes (see the attached movie Twisted-Filament.mp4). 

\begin{figure*}
\centering
\mbox{
 \includegraphics[scale=0.25]{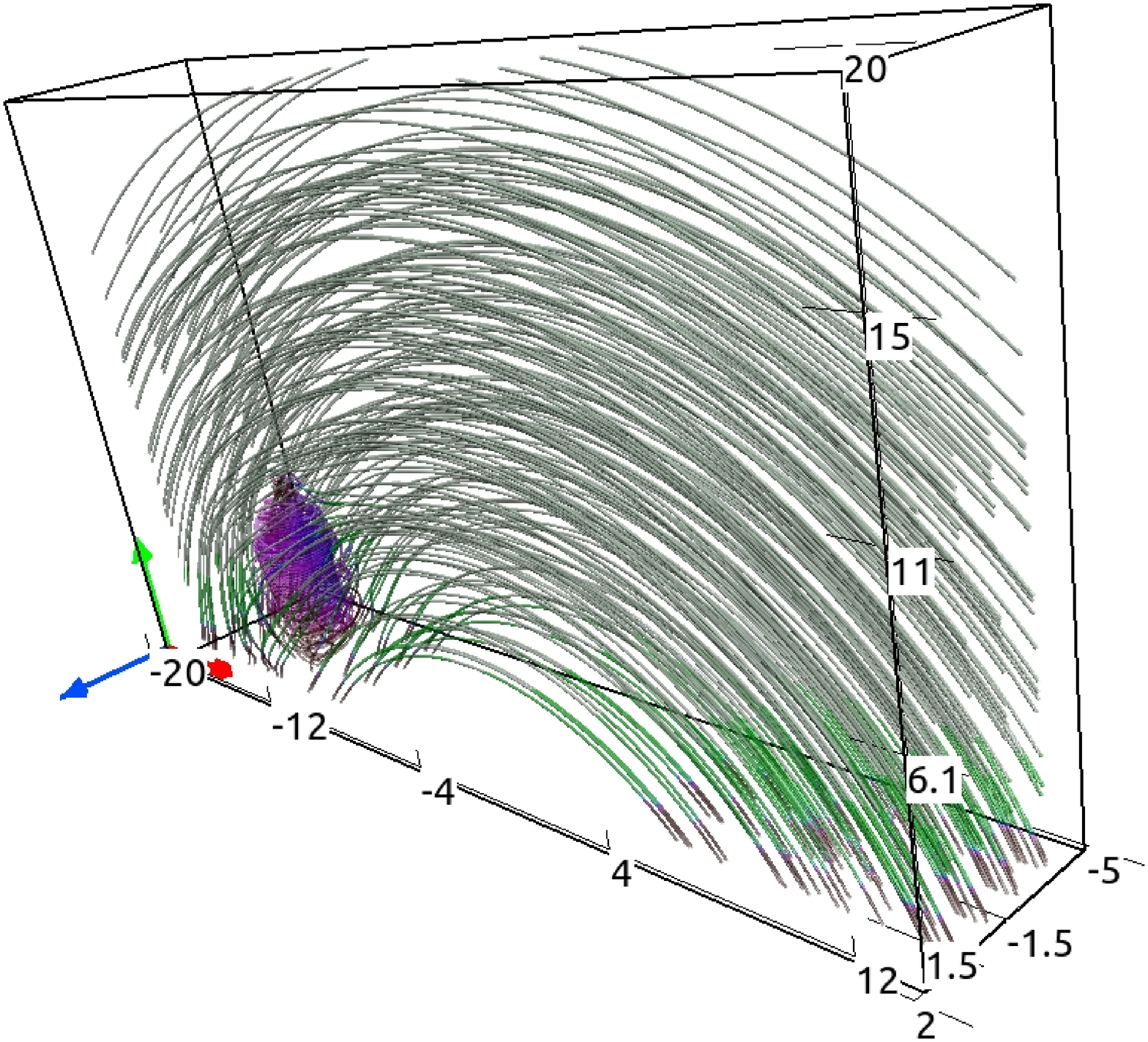}
\hspace{+1.0cm}
\includegraphics[scale=0.25]{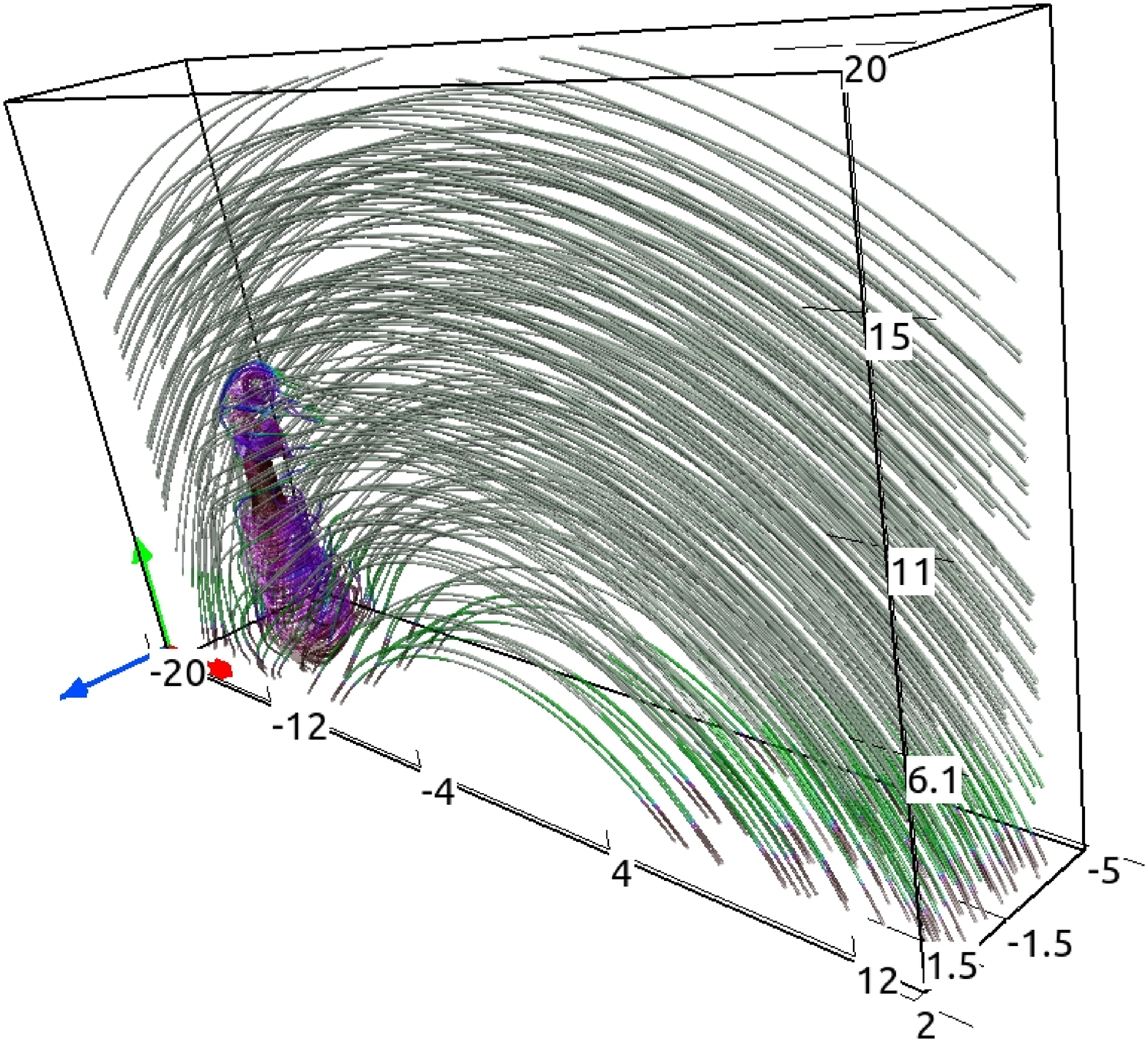}
}
\mbox{
\includegraphics[scale=0.25]{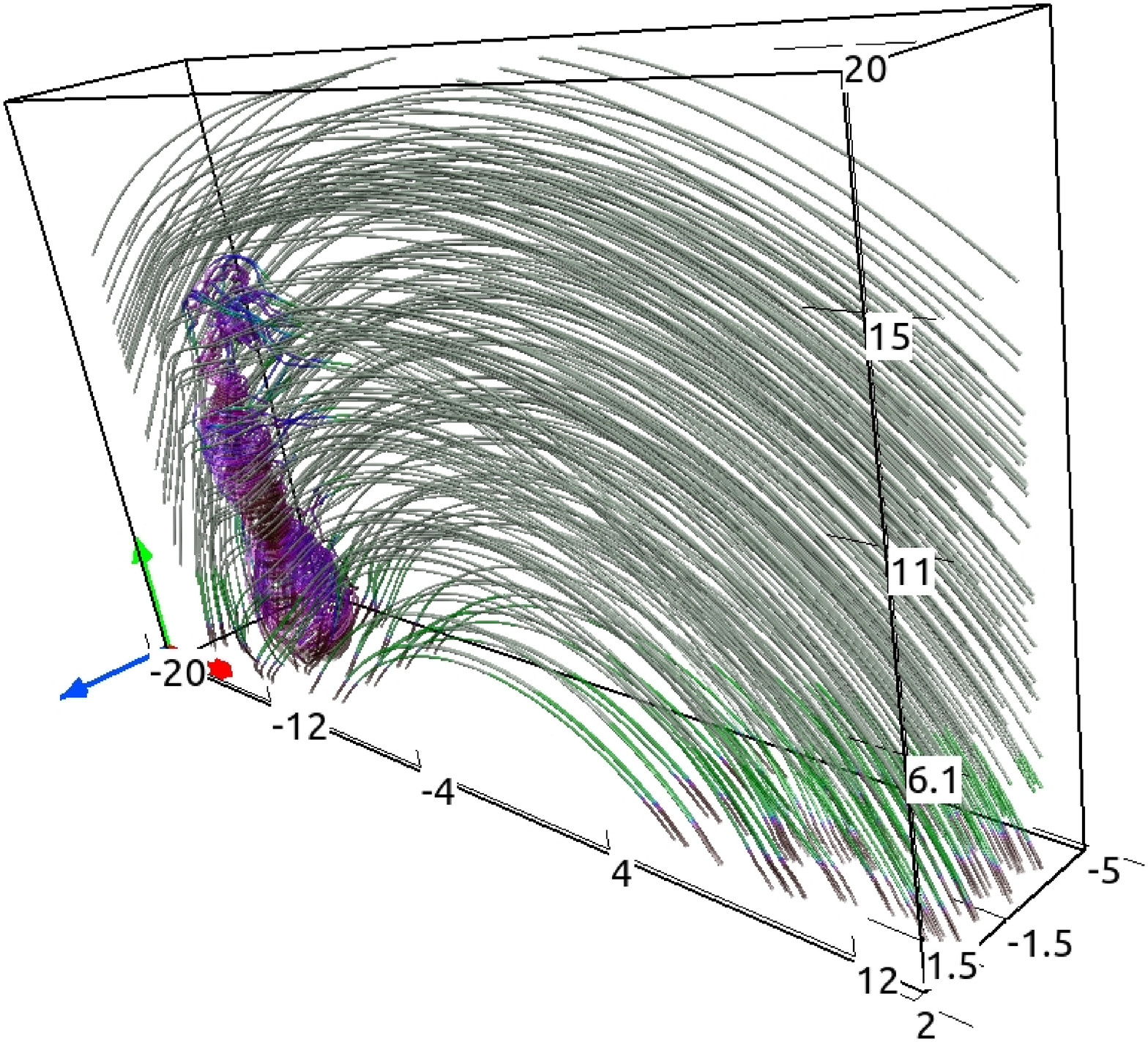}
\hspace{+1.0cm}
\includegraphics[scale=0.25]{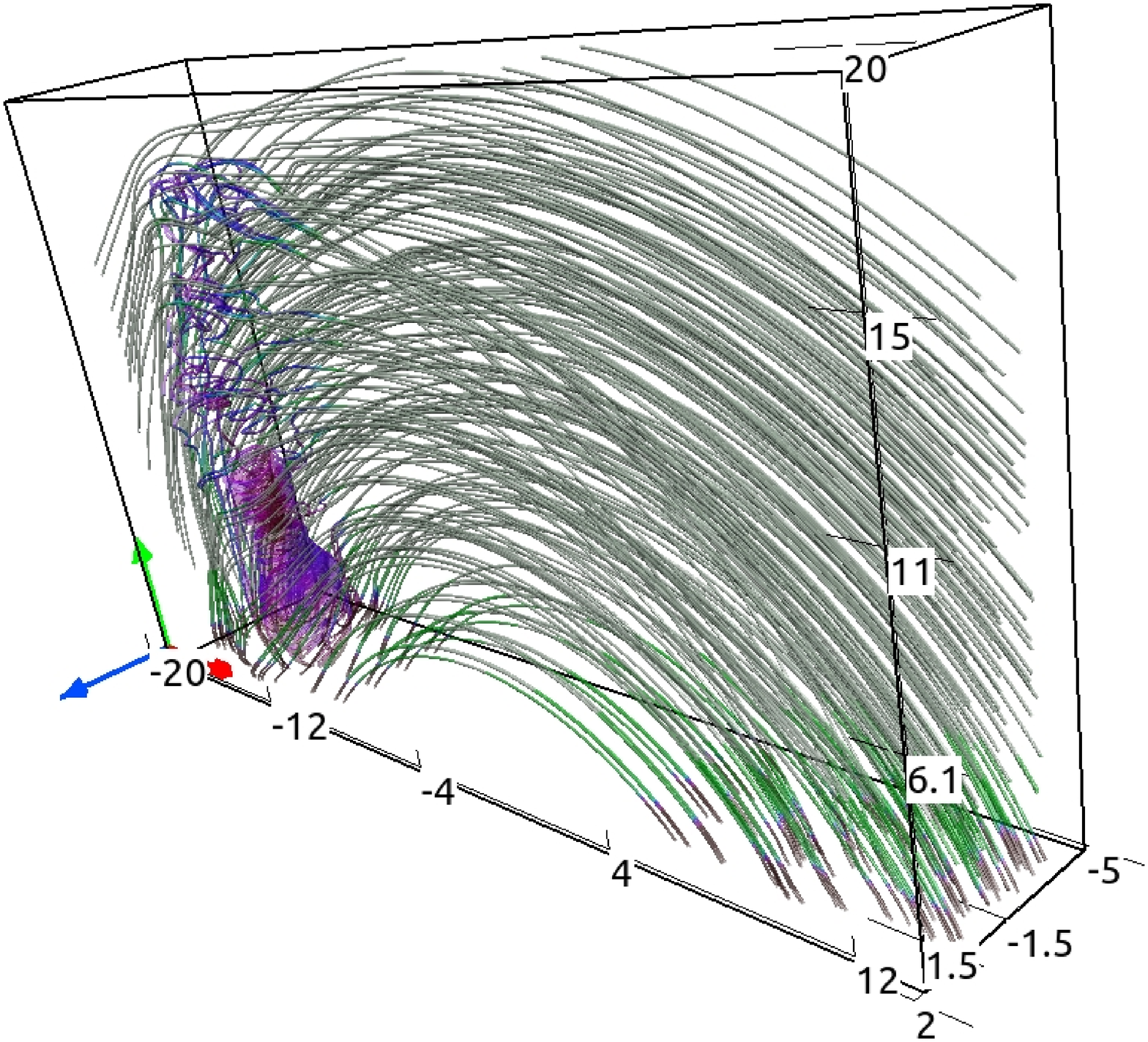}
}
\vspace{-1.0cm}
\caption{
Activation of right-handed clockwise twist
in the magnetic field of the flux-rope in corona 
at 
$t=50$ s (top-left), 
$t=75$ s (top-right), 
$t=100$ s (bottom-left), $t=125$ s (bottom-right). 
Red, green, and blue arrows correspond to the $x$-, $y$-, and $z$-axis, respectively.
}
\label{fig:B_lines_w}
\end{figure*}

\begin{figure*}
\centering
\mbox{
 \includegraphics[scale=0.25]{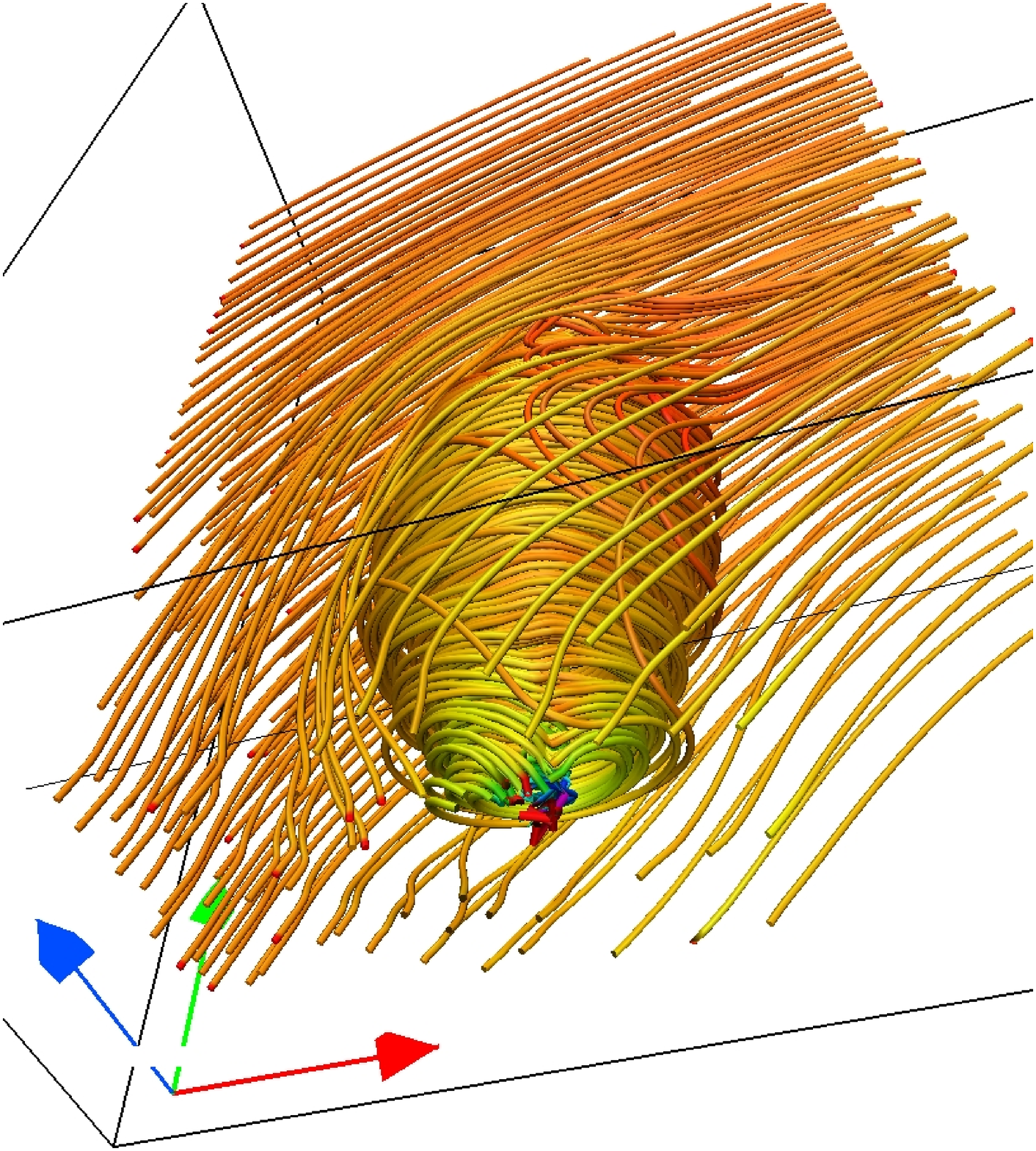}
\hspace{+1.0cm}
\includegraphics[scale=0.25]{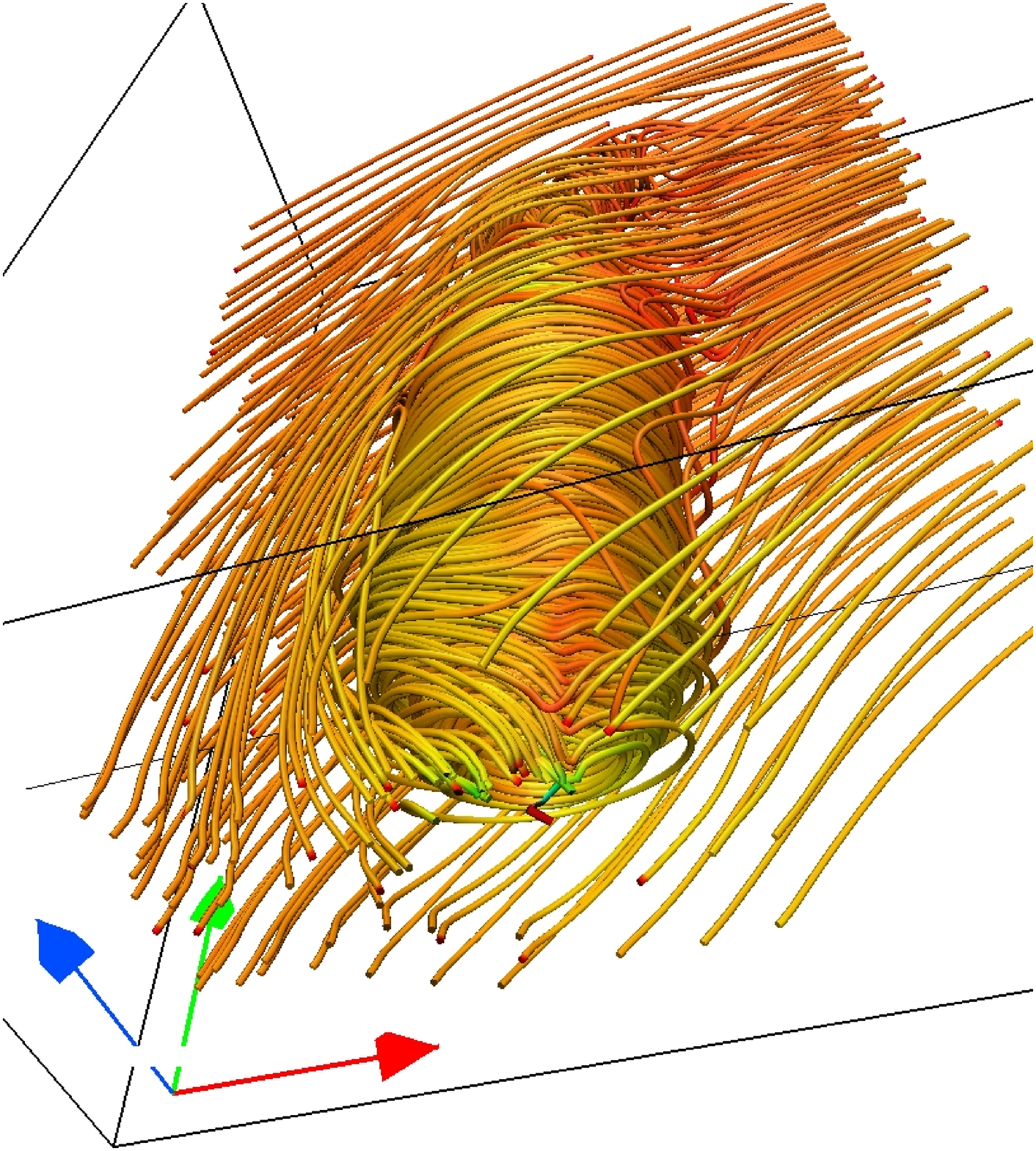}
}

\mbox{
\includegraphics[scale=0.25]{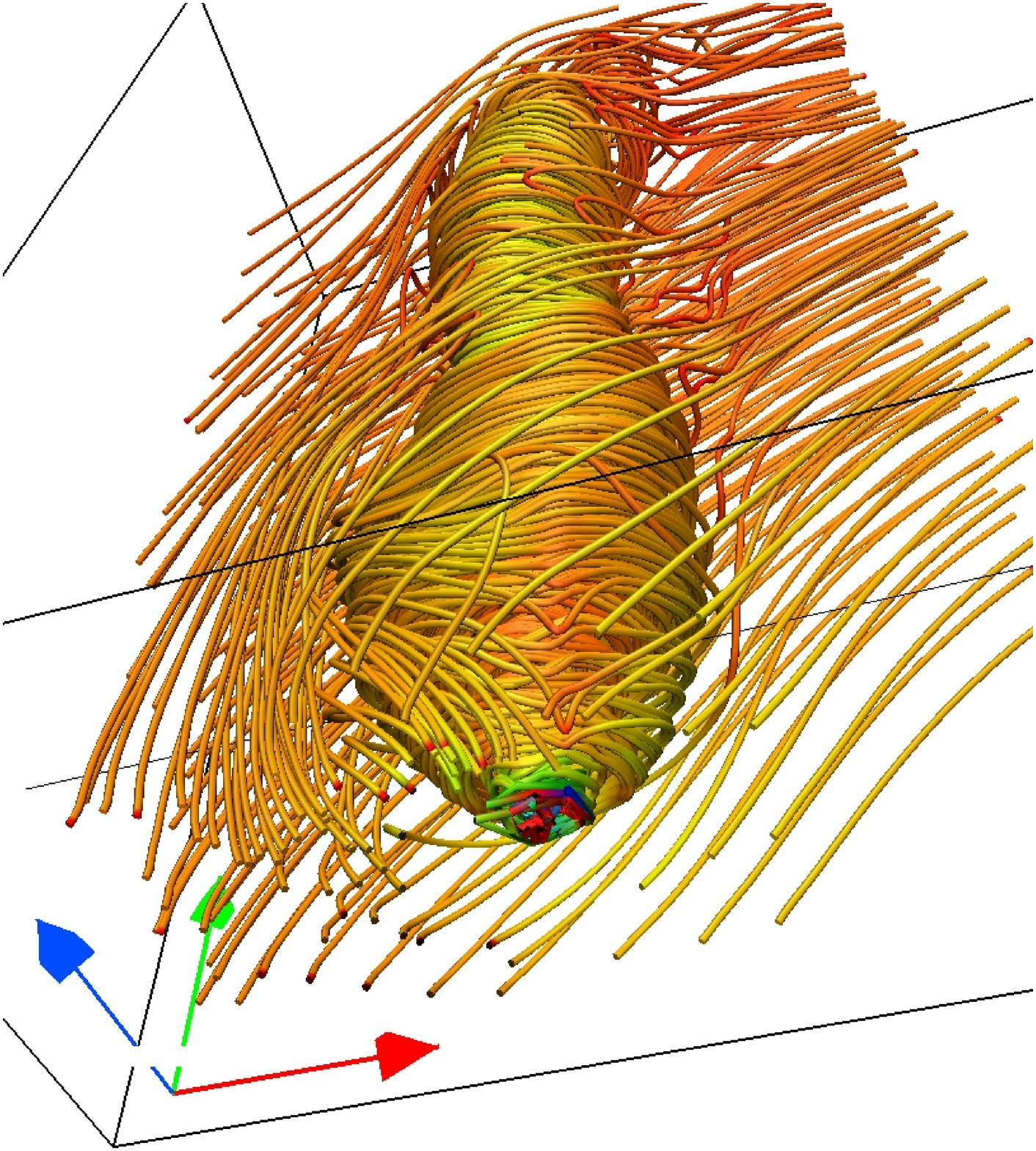}
\hspace{+1.0cm}
\includegraphics[scale=0.25]{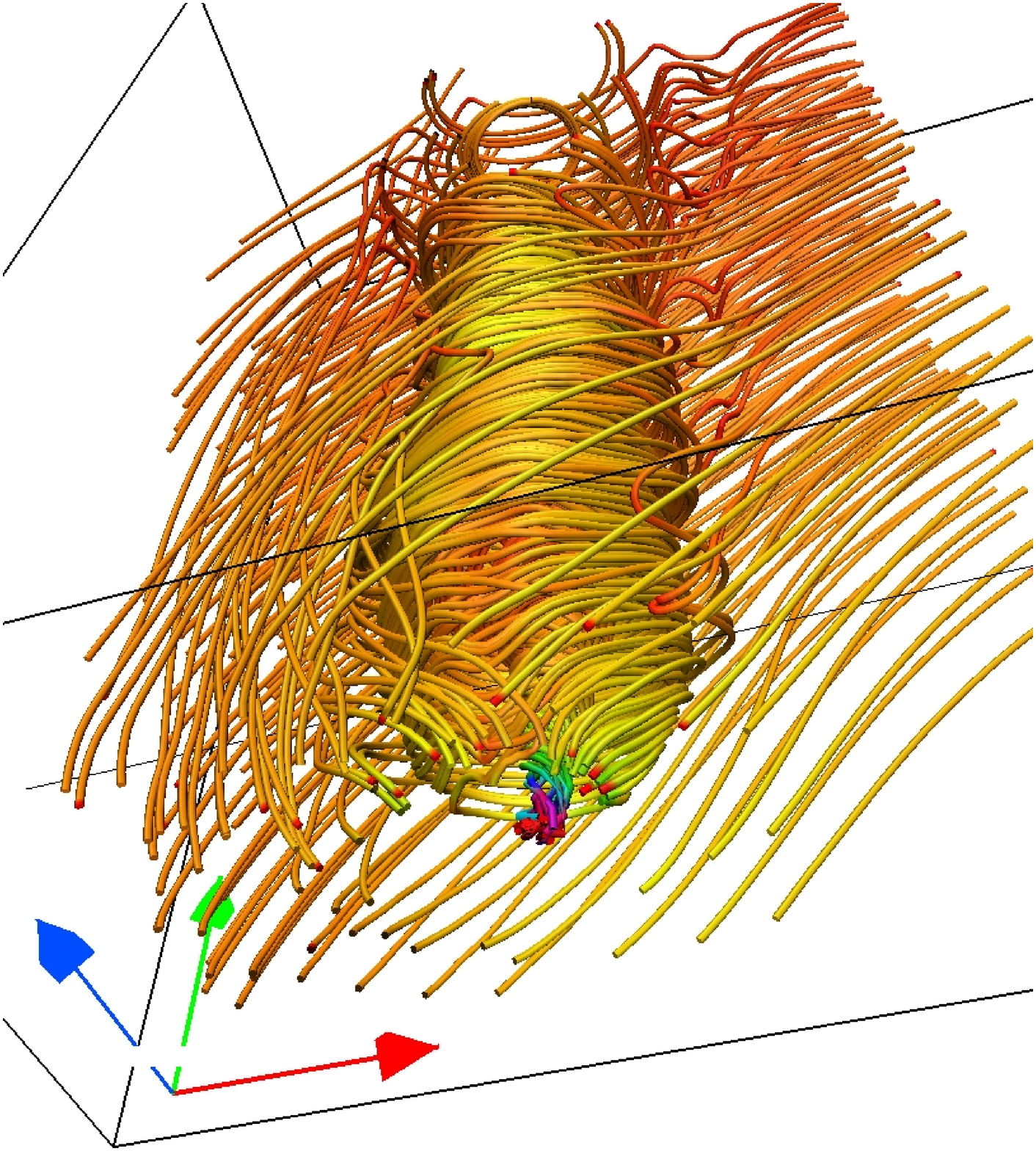}
}
\caption{Activation of right-handed clockwise twist
in the magnetic field of the flux-rope seen from near the photosphere 
at 
$t=50$ s (top-left), 
$t=75$ s (top-right), 
$t=100$ s (bottom-left), $t=125$ s (bottom-right). 
}
\label{fig:B_lines}
\end{figure*}

\begin{figure*}
\centering
\mbox{
\includegraphics[scale=0.25]{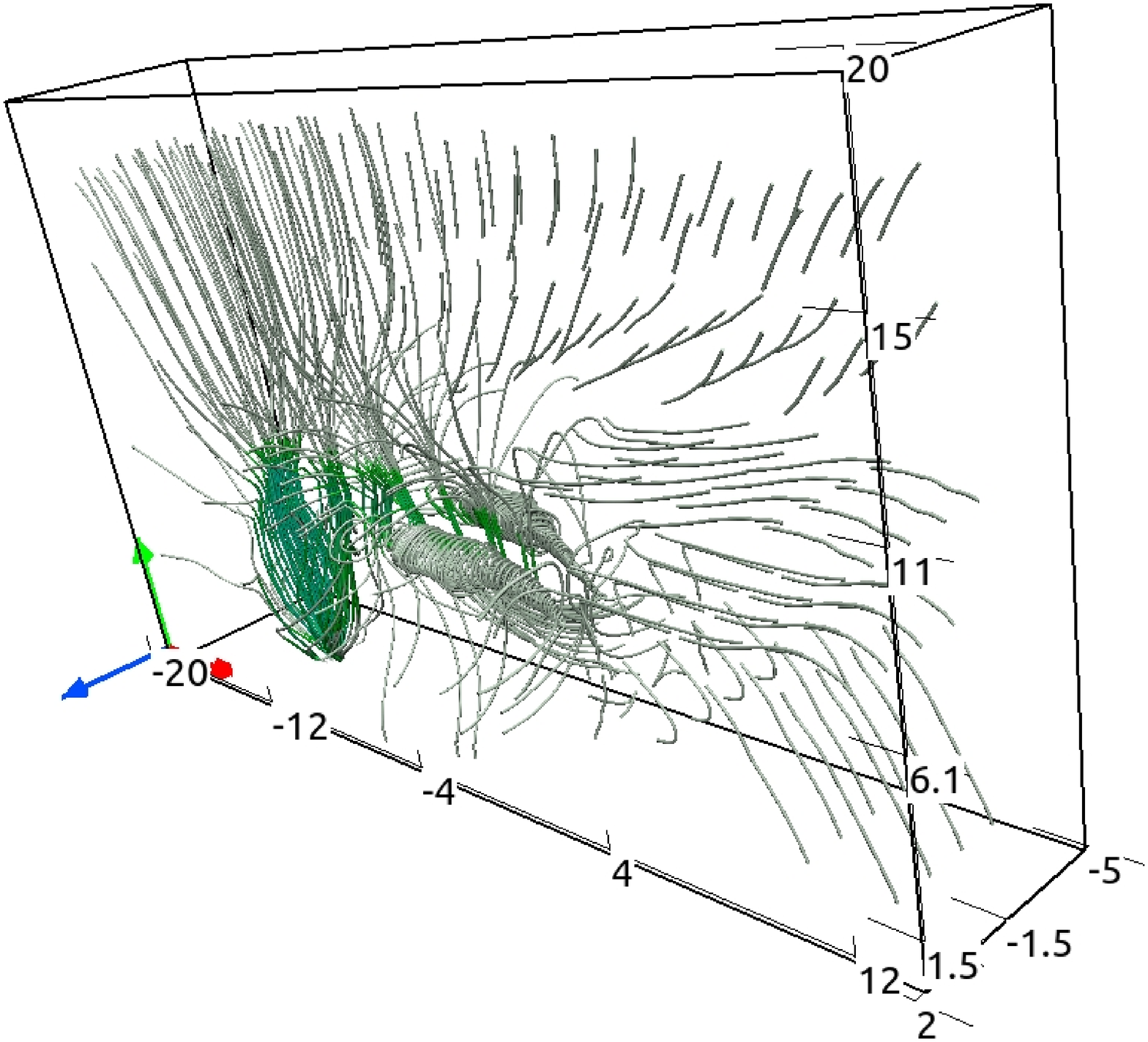}
\hspace{+1.0cm}
\includegraphics[scale=0.25]{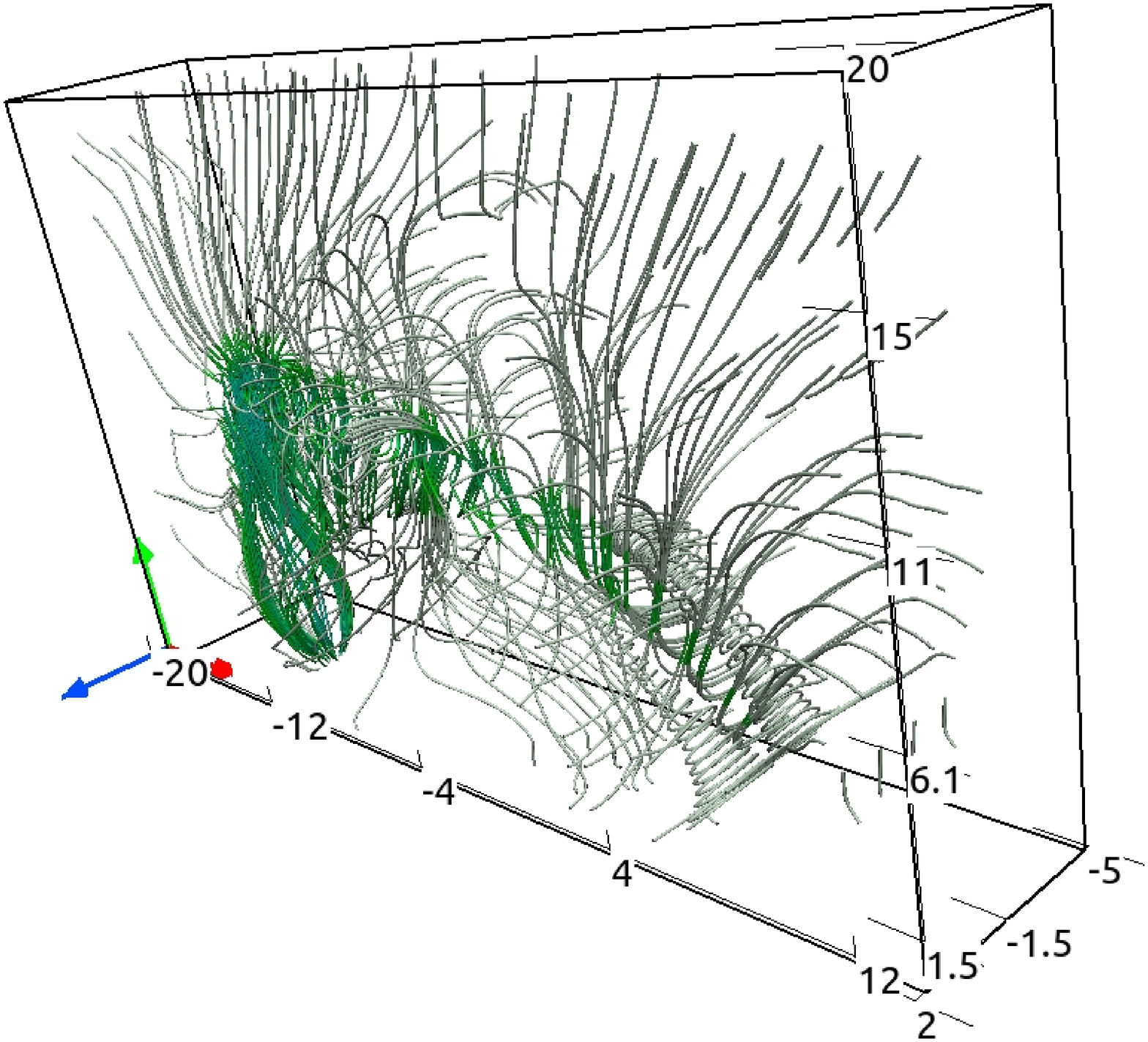}
}
\mbox{
\includegraphics[scale=0.25]{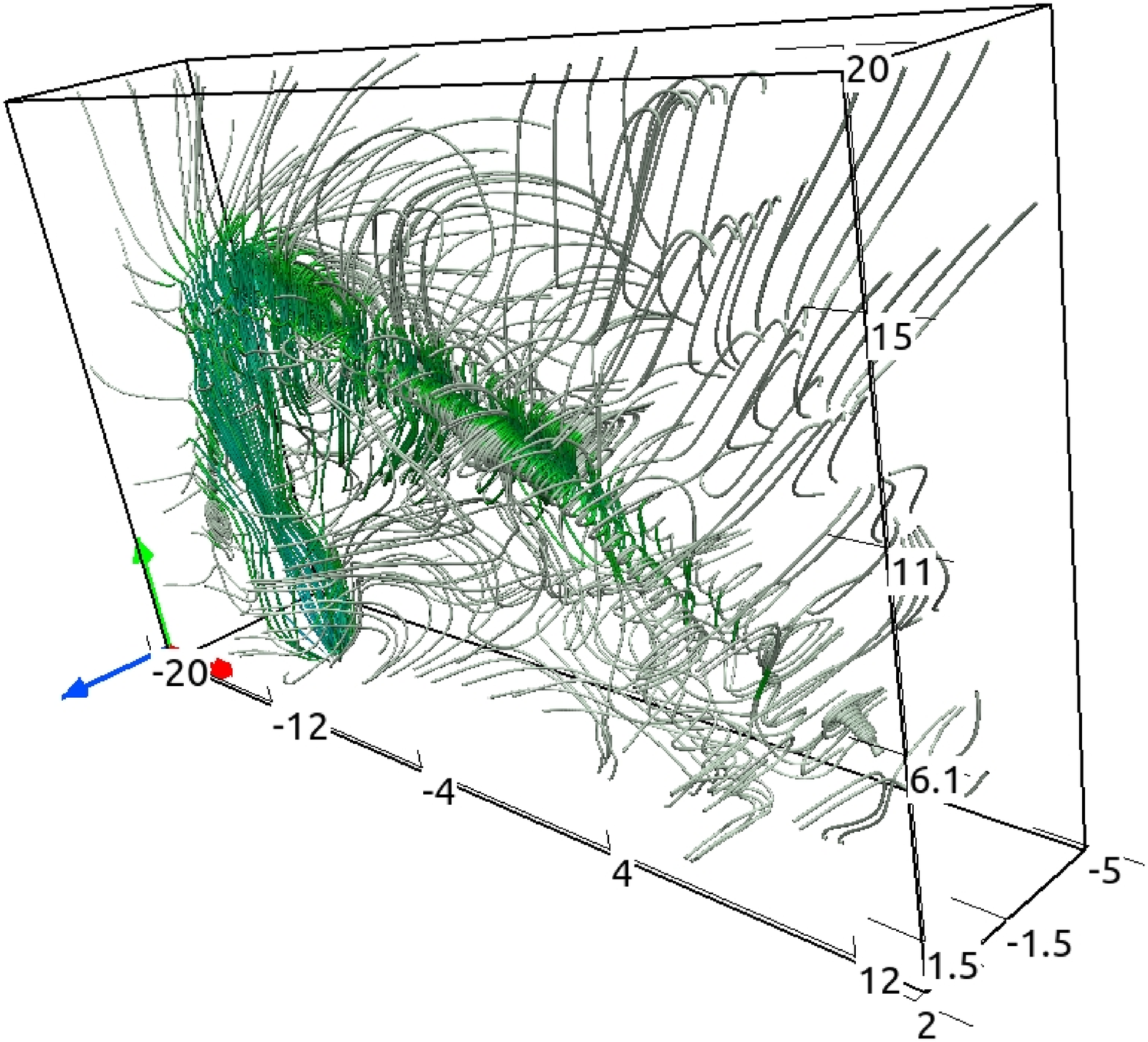}
\hspace{+1.0cm}
\includegraphics[scale=0.25]{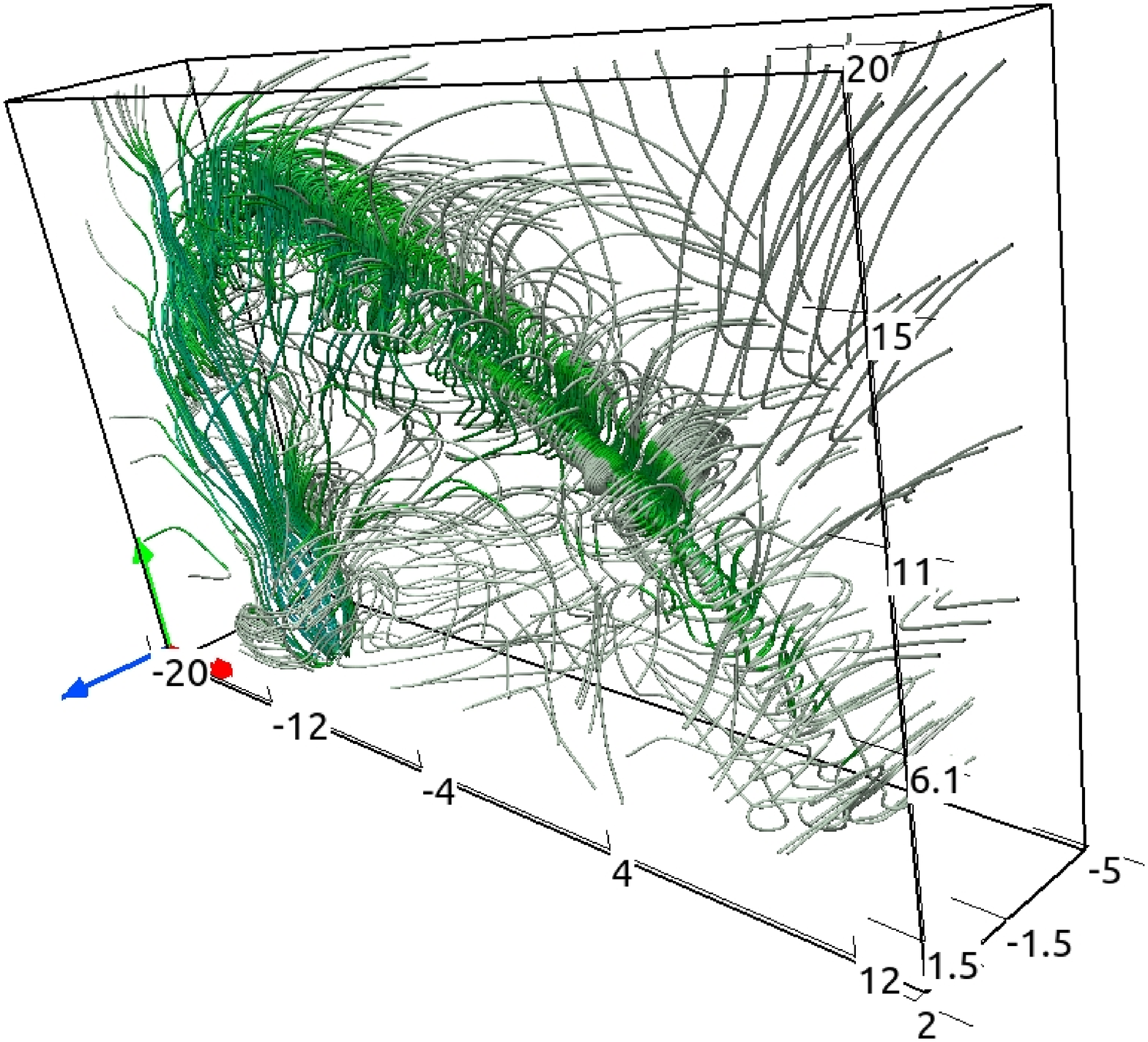}
}
\vspace{-1.5cm}
\caption{
Streamlines at 
$t=50$ s (top-left), 
$t=75$ s (top-right), 
$t=100$ s (bottom-left), and $t=125$ s (bottom-right). 
}
\label{fig:V-lines}
\end{figure*}

\begin{figure*}
\centering
\mbox{
\hspace{-2.0cm}
\includegraphics[scale=0.3, angle=90]{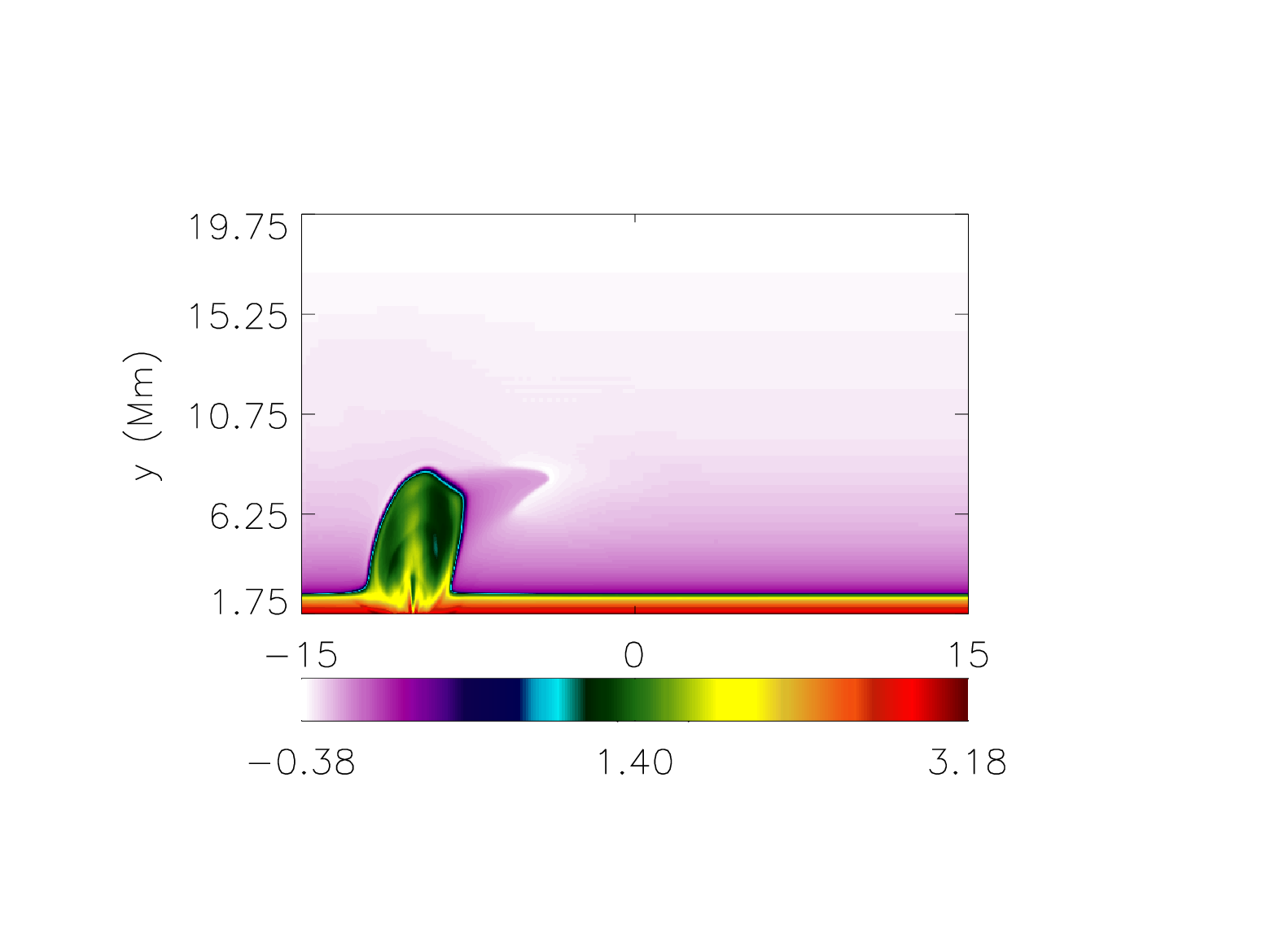}
\hspace{-2.0cm}
\includegraphics[scale=0.3, angle=90]{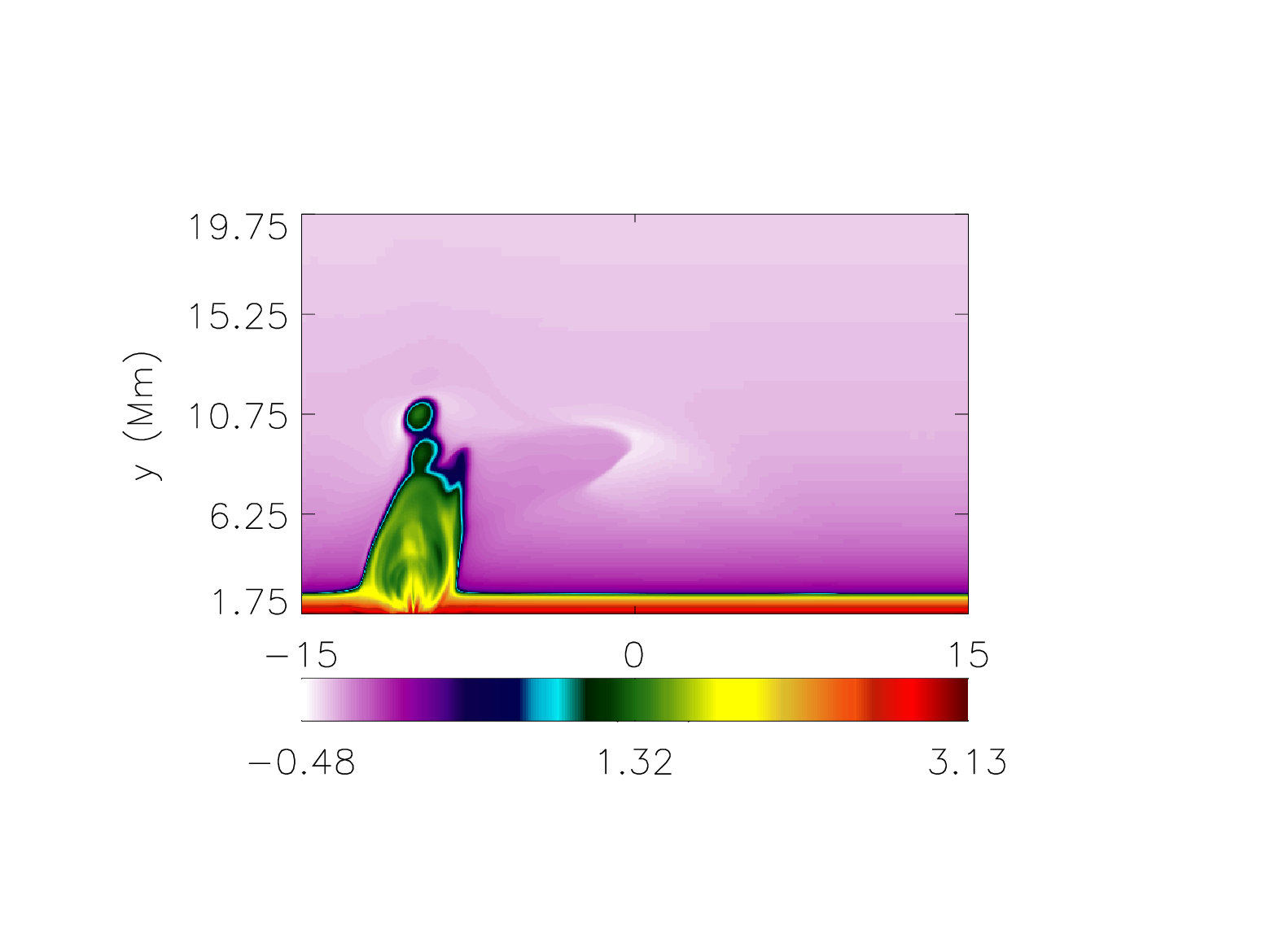}
}
\mbox{
\hspace{-2.0cm}
\includegraphics[scale=0.3, angle=90]{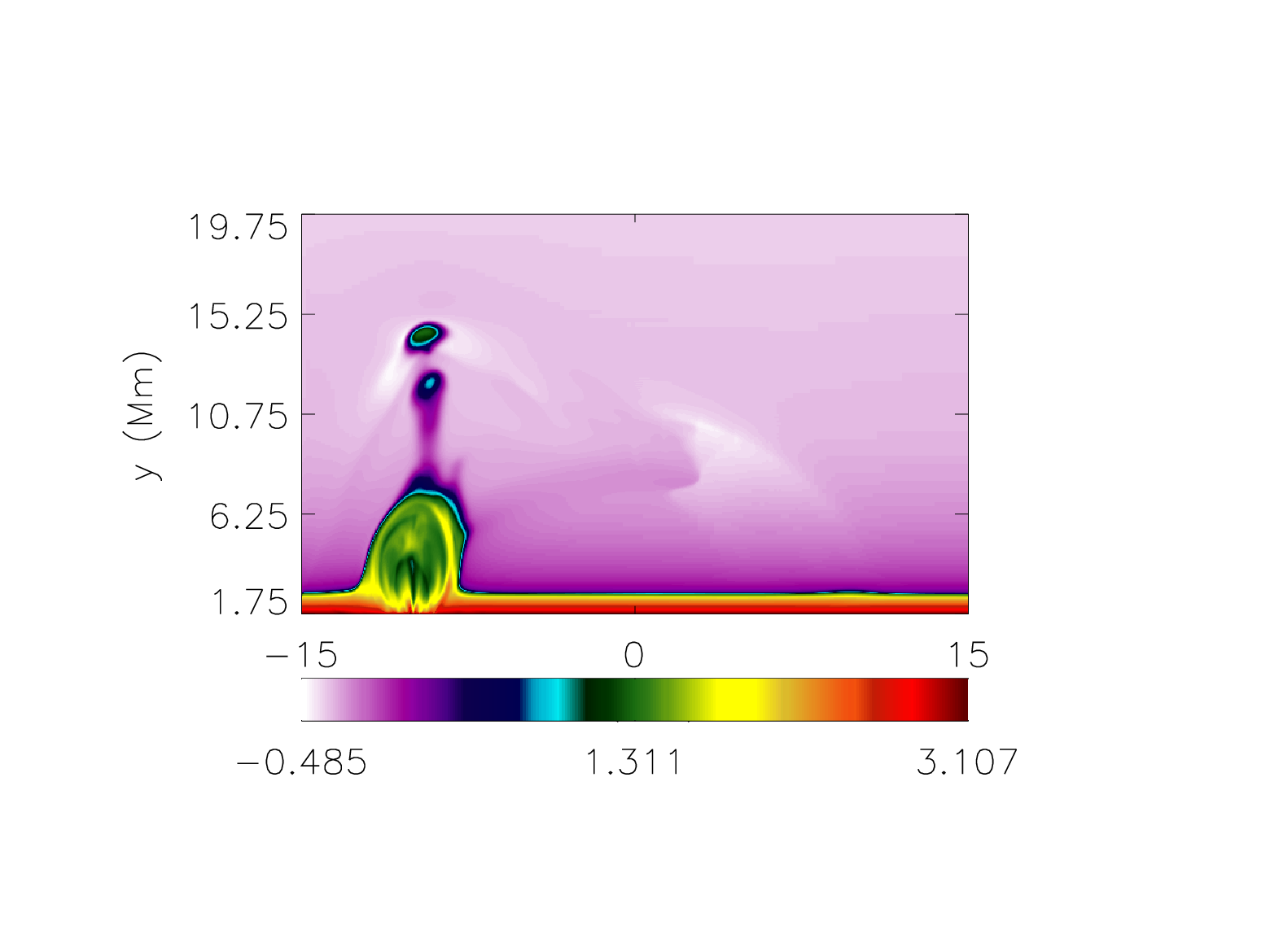}
\hspace{-2.0cm}
\includegraphics[scale=0.3, angle=90]{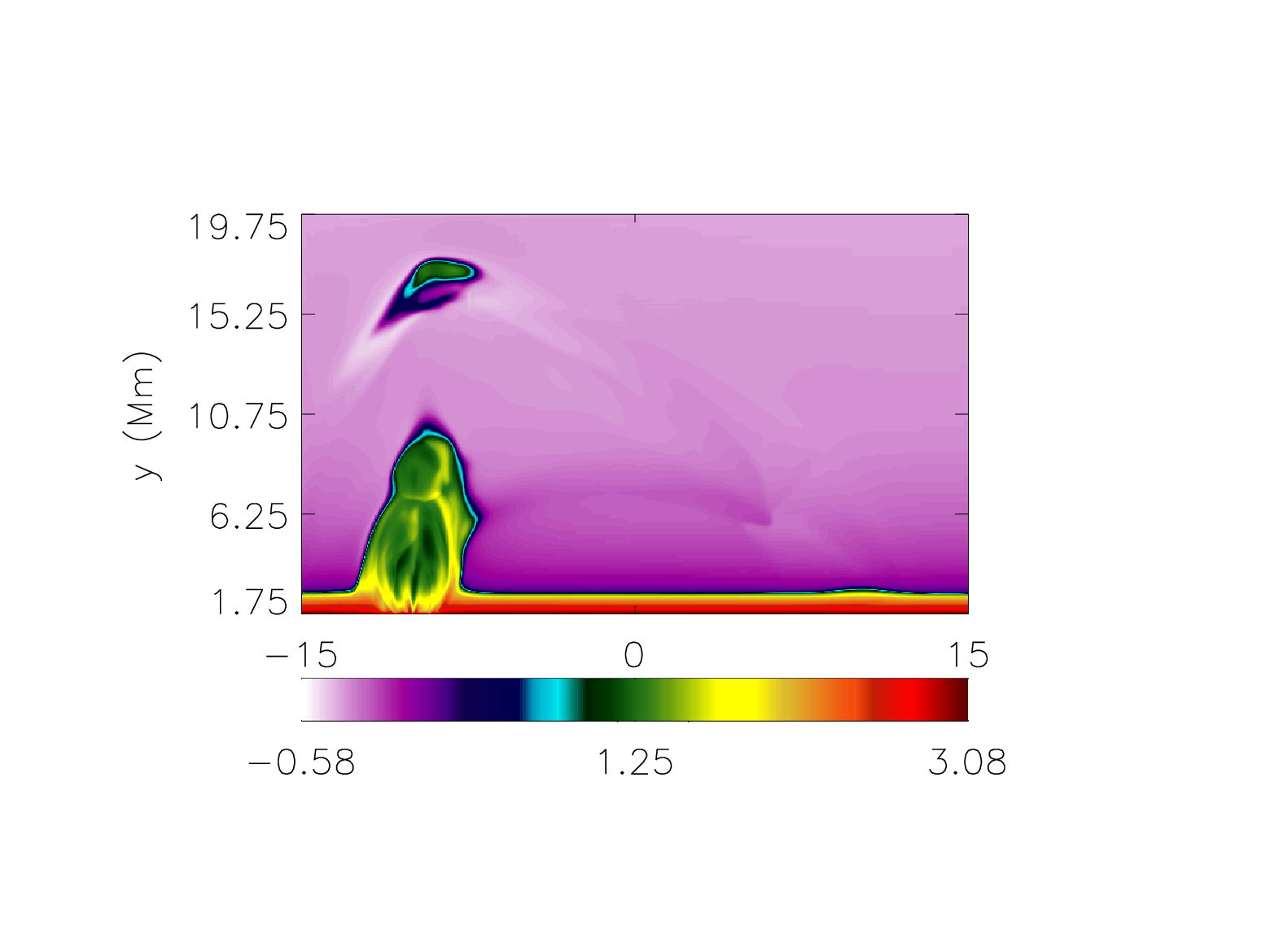}
}
\vspace{-1.5cm}
\caption{The mass density maps in X-Y plane (for $z=0$) 
at times $t=50$ s, $t=75$ s, $t=100$ s, and $t=125$ s. 
The evolution of shock front along the flux-rope field lines and the complex plasma motions are evident.
}
\label{fig:log_rho}
\end{figure*}

\begin{figure*}
\centering
\hspace{-2.0cm}
\includegraphics[scale=0.4]{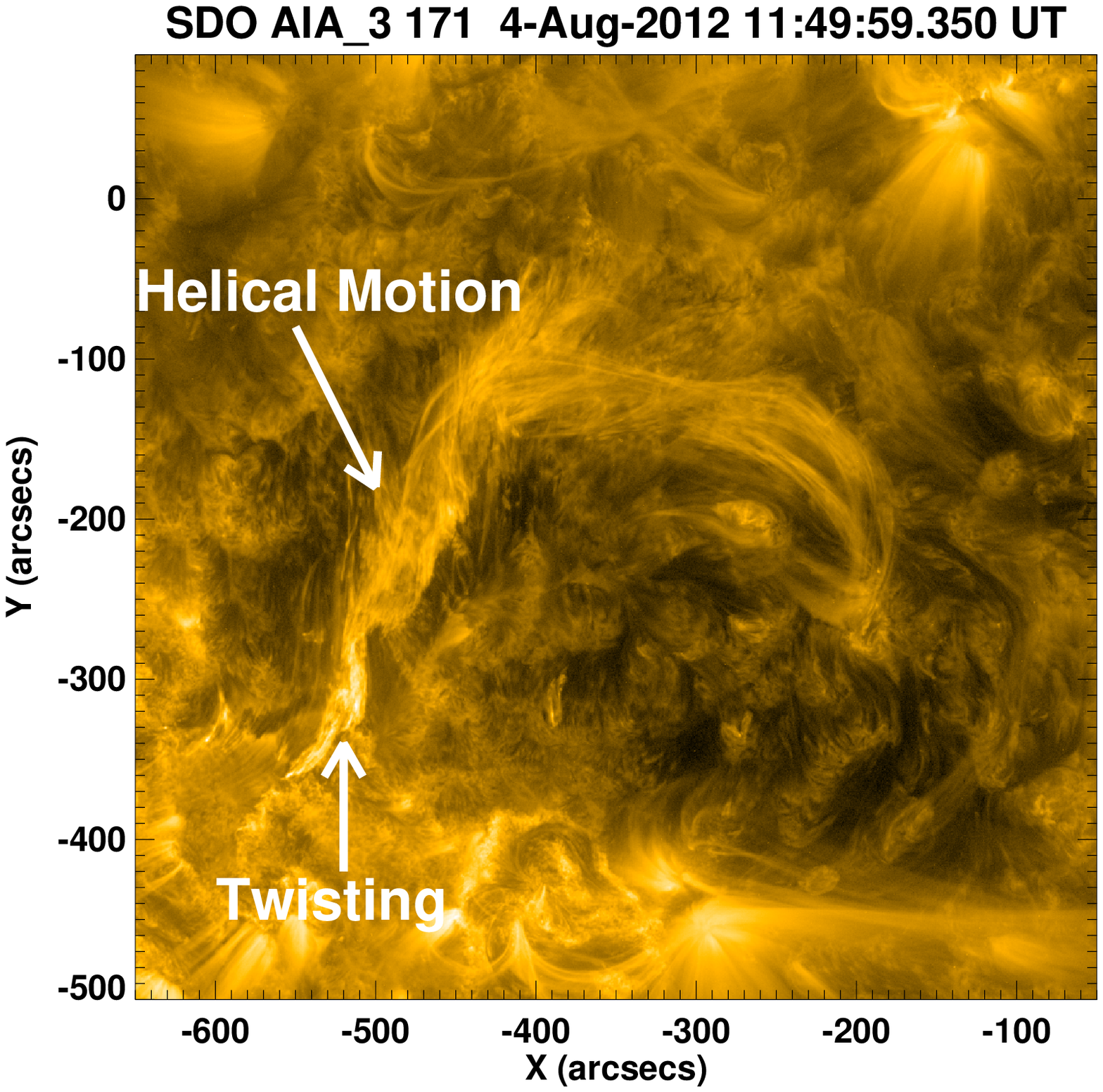}
\mbox{
\vspace{+5.0cm}
\hspace{-2.60cm}
\includegraphics[scale=0.4]{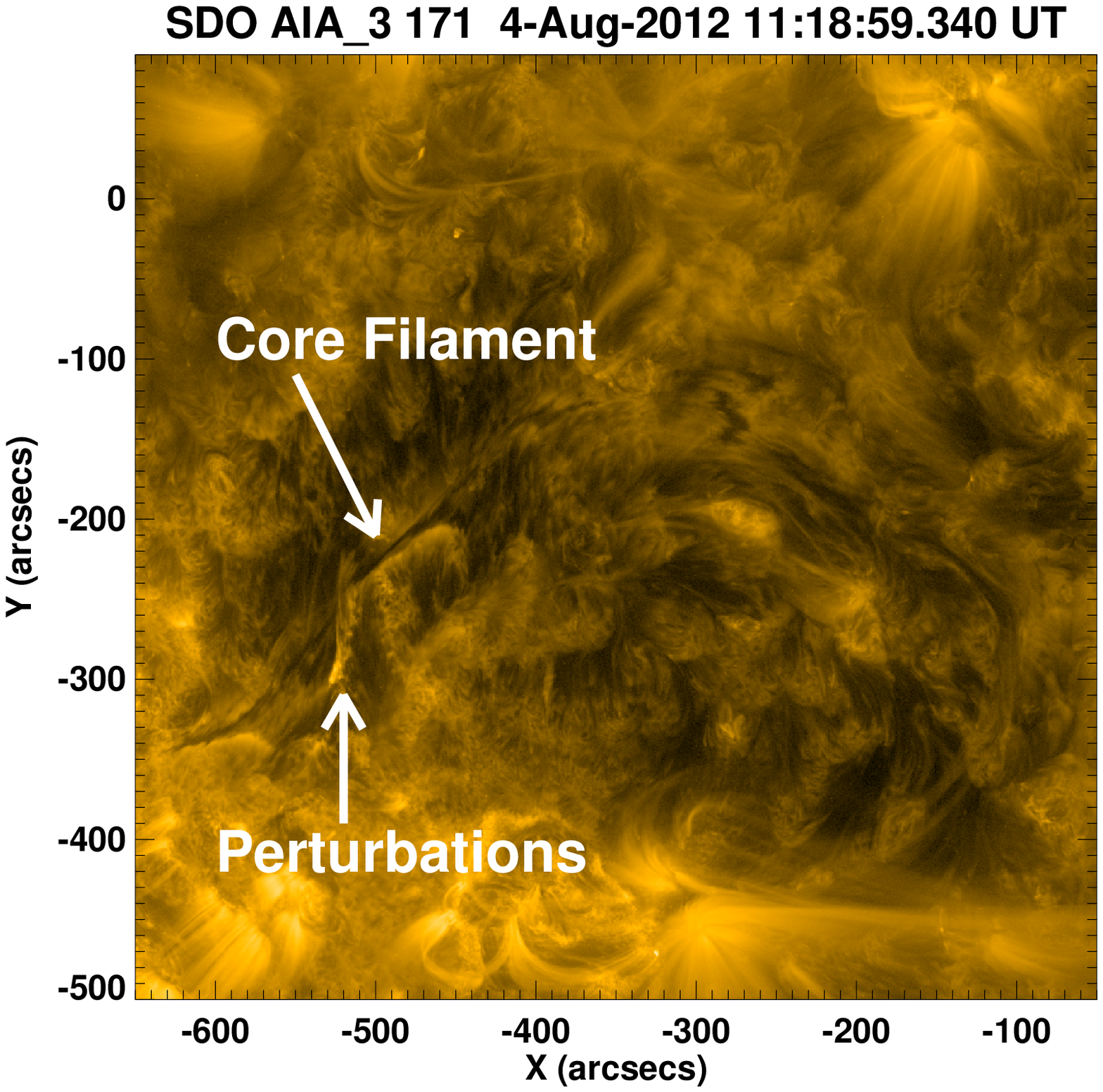}
\vspace{+5.0cm}
\hspace{-2.60cm}
\includegraphics[scale=0.4]{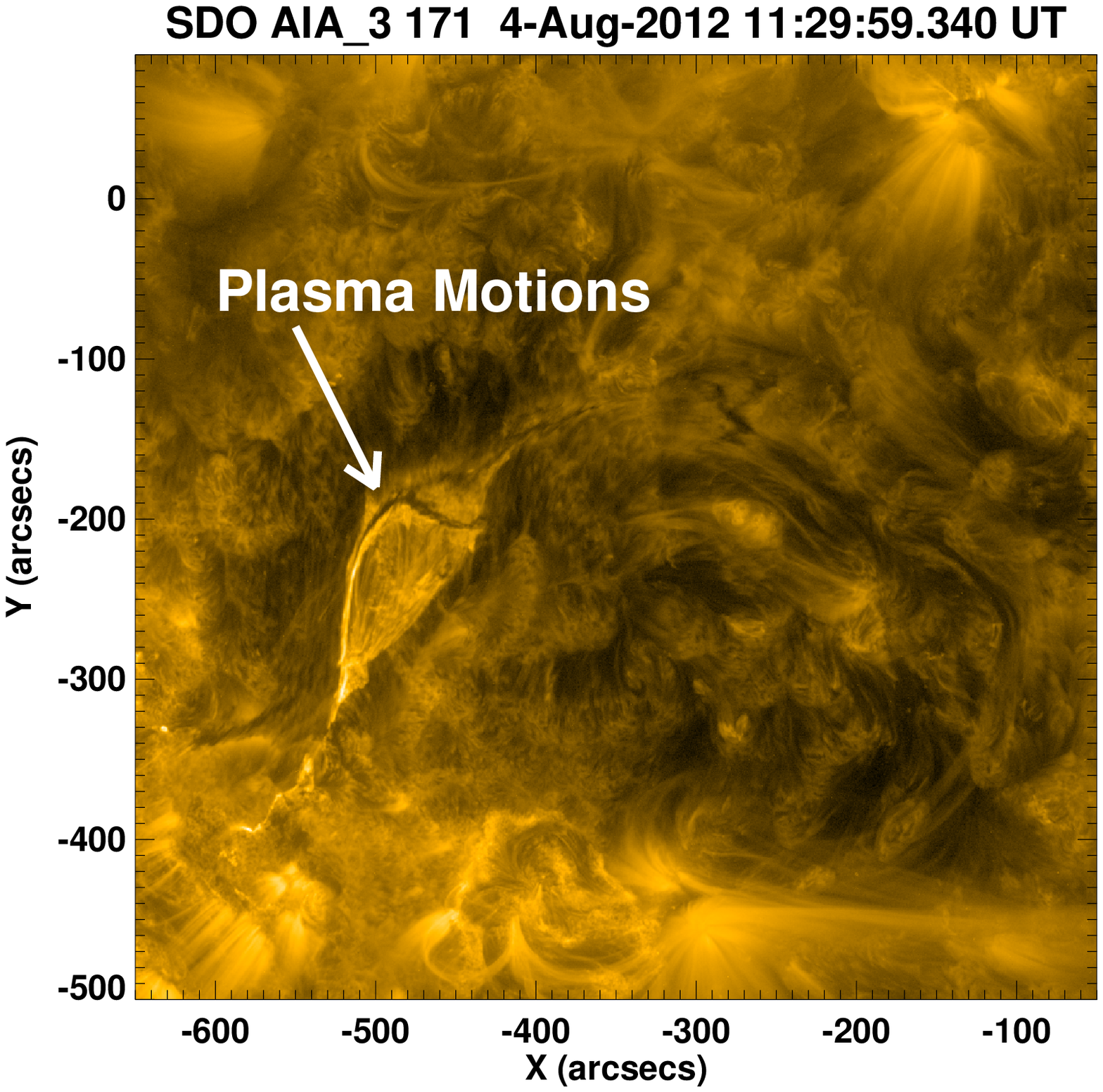}
\vspace{+5.0cm}
\hspace{-2.60cm}
\includegraphics[scale=0.4]{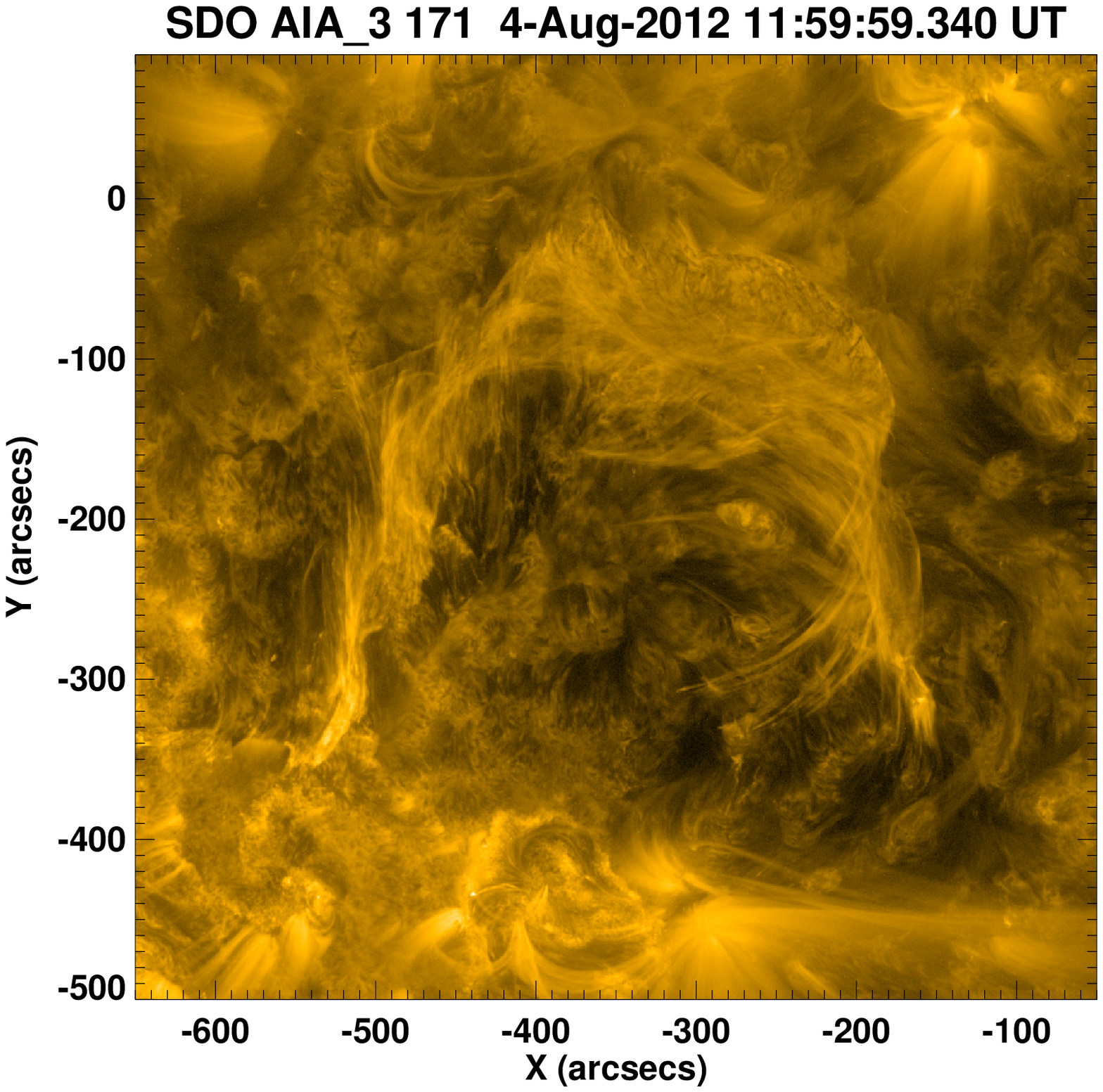}
}
\caption{Qualitative description of an activation of similar helical twists
and associated mass motion around a bipolar filament system though at
large spatio-temporal scale adopted from Joshi et al. (2014). 
}
\label{fig:obser}
\end{figure*}

Since we view the plasma motions in a filament plane that is 
almost perpendicular to the LOS, we compare the results of our numerical simulations 
(see Fig.~\ref{fig:B_lines_w}) to the observations (see Fig.~\ref{fig:obser} 
and the movie). As already mentioned above, the comparison can only be
qualitative because of the differences in scales between the numerical results
and the data; our simulation domain must be re-scaled by factor of $24$ in 
the flux-tube length, by $10$ in the width (or spread of perturbations) near 
the apex, and by $10$ in the time-scale in order to match the data. The 
comparison shows that the numerical results qualitatively match the observed 
magnetic field and plasma dynamics of the filament. For example, let us 
compare the 11:29 UT image in Fig.~\ref{fig:obser} to the top-left snapshots 
of Figs.~\ref{fig:B_lines_w}-\ref{fig:B_lines}. Then, the activation of 
magnetic twists (black filament threads in Fig.~\ref{fig:obser} and twists 
in the magnetic field lines (magenta colors) in Fig.~\ref{fig:B_lines_w}) 
can clearly be seen. An initiation of the bright plasma swirl/vortex motions 
(the bright right-handed turning of the plasma envelope as shown in 
Fig.~\ref{fig:obser} and vortex in the magnetic field lines (magenta colors) 
in Fig.~\ref{fig:B_lines_w}) also match with each other. 

At the later times, the magnetic twists as well as the vortex motion of the 
plasma are evident both in the numerical simulations as well as in the 
observations. The fragmentation can be seen near the apex of the prominence 
fine structured plasma motions, where the formation of the two vortex channels 
in the numerical domain (cf., Fig.~\ref{fig:obser}, bottom-right and Fig.~5, bottom) 
takes place. 
Now, the process of formation of the fragmented two vortex channels becomes evident in our numerical simulations 
(shape of eight in Figs.~\ref{fig:B_lines} and 5)
than in the observations. The reason is
that the real twist in the observational regime may differ from that considered 
in our model. Moreover, as stated above, the fragmentation of the vortex/swirls 
of the plasma along the two channels can also be seen in Fig.~\ref{fig:obser}.
It must be noted nothing that both the observed and simulated vortex channels 
are associated with fast speeds that are the signatures of the evolution of 
the torsional Alfv\'en wave perturbations moving through these channels 
in a large-scale prominence-like bipolar magnetic structure. 
We referred to the observed and 
simulated phenomena as the fast magnetic twisters associated with the torsional perturbations 
and fast vortex motions.
\section{Discussion and Concluding Remarks}
We investigated numerically physical implications of the activation 
of the magnetic twists in a potential coronal magnetic flux-tube embedded 
in the solar atmosphere with a realistic temperature distribution. Our 
numerical results reveal the evolution of right-handed magnetic twists, 
double vortex channels, fragmentation and fast propagating perturbations, 
all evident in our coronal arcade model in which variable twists in the 
azimuthal component of the magnetic field were initially imposed. The 
result is a fast magnetic twister whose existence is reported here for 
the first time. 

The initial perturbations imposed in our coronal arcade model generate 
torsional Alfv\'en waves as well as plasma swirls that reach the other 
foot-point of the arcade and partially reflect back from the transition region. 
The two vortex channels are evident in the generated twisted flux-tube with 
a fragmentation near its apex that results from the initial twist as well as 
from the morphology of the tube. This highly depends upon the initial 
magnetic field configuration, plasma properties, and nature of perturbations, 
which all determine how the vortices, associated waves and plasma motions are 
formed.

There were some previous studies of vortex motions at various 
spatio-temporal scales (e.g., Bonet et al. 2008; Shelyag et al. 2011; 
however, in none of them the fast magnetic twister was identified. 
Moreover, Shelyag et al. (2013) reported no vortices (tornado-like motions) 
in their numerical simulations but only torsional Alfv\'en waves. We do 
see both fast magnetic twisters and torsional Alfv\'en waves. The difference 
between their approach and ours is in temporal scales, which in their approach 
are much shorter than in ours. Based on our results, we conclude that a tornado 
needs a long lasting twist in order to be sustained for time scales of the order 
of 12 hours or longer. 

Our numerical results were compared to the observational data of plasma 
motions in a solar prominence, whose rare observation showed the evidence 
for the existence of a fast twister. Interestingly enough, the general 
properties of this twister are similar to those seen in our numerical 
simulations. However, we could only show that are numerical results and 
the data agreed only qualitatively because significant differences in the 
spatio-temporal scales between the observations and our numerical simulations
prevented us from making a more direct comparison. In the future, more 
observations are needed to establish validity of our coronal arcade model
at various spatio-temporal scales, and to examine if the double vortex 
system, plasma swirling, and fast torsional perturbations, all exist
in a single prominence system before its eruption.
\clearpage
\section{Acknowledgments}
The authors are indebted to an anonymous referee whose valuable 
comments and suggestions allowed us to significantly improve the paper.
This work was supported by NSF under the grant AGS 1246074 (K.M. and Z.E.M.). 
A.K.S. acknowledges a support during his stay at UMCS in Lublin, Poland,
when a significant portion of this work was done. He also thanks Dr. N.C. 
Joshi for providing the observational data and animation, as well as Shobhna for 
patient encouragements. 
Z.E.M. acknowledges the support of this work by 
the Alexander von Humboldt Foundation, 
and by University of Texas at Arlington through its Faculty Development Program. 
We also acknowledge the use of the 
SDO/AIA observations for this study, with the data provided courtesy of 
NASA/SDO, LMSAL, and the AIA, EVE, and HMI science teams. The FLASH code 
used in our numerical simulations was developed by the DOE-supported 
ASC/Alliance Center for Astrophysical Thermonuclear Flashes at the 
University of Chicago. 
A.K.S. (and other co-authors) dedicate this work to one of the pioneering 
plasma physicists Prof. P.K. Shukla (7 July, 1950 – 26 January, 2013), 
Ruhr-University Bochum (RUB), Germany, who also had a doctoral degree in 
Department of Applied Physics, I.T.BHU, Varanasi, India in 1972, where he 
also completed his doctoral research in 2006. 
{}
%
%

%
\clearpage
%

%
\clearpage
%


%
\clearpage
%


%
\clearpage
%
%
\clearpage
%

\clearpage

\clearpage

\clearpage

\end{document}